\RequirePackage{fix-cm}
\documentclass[twocolumn,epjc3]{svjour3}  
\smartqed  
\RequirePackage{graphicx}
\RequirePackage{amssymb}
\RequirePackage{lipsum}
\RequirePackage{amsmath}
\usepackage[normalem]{ulem}
\usepackage{appendix}
\RequirePackage{simpler-wick}
\RequirePackage{slashed}

\newcommand{\widesim}[2][1.5]{
  \mathrel{\overset{#2}{\scalebox{#1}[1]{$\sim$}}}
}
\journalname{Eur. Phys. J. C}
\begin{document}

\title{Heavy-quark mass effects in off-light-cone distributions}


\author{Valerio Bertone\thanksref{e1,addr1} \and Michael Fucilla\thanksref{e2,addr2} \and Cédric Mezrag\thanksref{e3,addr1}
}

\thankstext{e1}{e-mail: valerio.bertone@cea.fr}
\thankstext{e2}{e-mail: michael.fucilla@ijclab.in2p3.fr}
\thankstext{e3}{e-mail: cedric.mezrag@cea.fr}


\institute{Irfu, CEA, Universite Paris-Saclay, Gif-sur-Yvette, F-91191, France \label{addr1}
           \and
Universite Paris-Saclay CNRS/IN2P3, IJCLab, Orsay, F-91405, France \label{addr2}         
}

\date{Received: date / Accepted: date}

\maketitle

\begin{abstract}
We compute the one-loop correction to the forward matrix element of
  an off-light-cone bi-local quark correlator characterised by a
  space-like separation $z^2$ in the presence of heavy quarks with
  mass $m$. This calculation allows us to extract the one-loop
  matching kernel, necessary to connect quasi- and pseudo-distributions
  to collinear parton distribution functions (PDFs), accounting for
  heavy-quark mass effects. Our result is exact in that it includes
  all powers of $z^2m^2$ at one loop. In the limit
  $z^2m^2\rightarrow 0$, it consistently reduces to the known massless
  result. We also carry out a numerical implementation of our expressions, which
  allows us to compute the charm pseudo-distribution of the proton
  given its PDFs. We finally comment on the quantitative impact of
  heavy-quark mass corrections.
\end{abstract}

\section{Introduction}
\label{introduction}

In the past decades, much effort has been put into attempting to
extract information on the structure of hadrons from lattice
simulations of Quantum Chromodynamics (QCD) (see \emph{e.g.} the
reviews in
Refs.~\cite{Cichy:2018mum,Radyushkin:2019mye,Constantinou:2020hdm}).
However, the task is complicated by the fact that most of the
phenomenologically relevant partonic distributions are defined through
bi-local partonic operators characterised by light-like separations.
Typical examples are parton distribution functions (PDFs) and
distribution amplitudes (DAs).

Because of the use of euclidean metric, light-like distances in
lattice-QCD simulations are reduced to a point, limiting the studies
to local operators related to moments of the distributions of
interest. Moreover, the breaking of Lorentz symmetry generates
complicated mixings between operators, effectively restricting the
computation to the lowest moments. In spite of early attempts to
overcome this issue~\cite{Liu:1993cv,Detmold:2005gg,Braun:2007wv}, the
breakthrough came in 2013 with Ref.~\cite{Ji:2013dva}, introducing the
Large-Momentum Effective Theory (LaMET) formalism, which for the first
time gave direct access to the momentum dependence of light-cone
distributions. The new formalism was followed by the so-called
short-distance factorisation approach~\cite{Radyushkin:2017cyf}, which
allows for a simpler connection between lattice simulations and
momentum dependence of light-cone distributions, through re\-nor\-ma\-li\-sa\-tion-group-invariant ratios.
Other formalisms have
also been developed (see, \textit{e.g.},
Refs.~\cite{Davoudi:2012ya,Chambers:2017dov,Ma:2017pxb,Bali:2018nde,Shindler:2023xpd}).

In both LaMET and short-distance factorisations, off-light-cone
distributions are related to light-cone distributions by means of
perturbative matching kernels. It is precisely these relations that
allow light-cone distributions, such as PDFs and DAs, to be extracted
from lattice simulations. Currently, LaMET matching kernels for PDFs
are known up to next-to-next-to leading
order~\cite{Xiong:2013bka,Ji:2014hxa,Ji:2015qla,Ji:2017rah,Radyushkin:2016hsy,Izubuchi:2018srq,Radyushkin:2018nbf,Li:2020xml,Chen:2020iqi,Chen:2020ody},
\textit{i.e.}  $\mathcal{O}(\alpha_s^2)$ in the QCD strong
coupling. Recently, the first three-loop (N$^3$LO) calculation for  unpolarised flavour non-singlet distributions has been achieved~\cite{Cheng:2024wyu}.
In the short-distance-factorisation formalism, instead, they
are available up to one
loop~\cite{Radyushkin:2017cyf,Radyushkin:2018cvn,Radyushkin:2017lvu,Radyushkin:2019owq,Balitsky:2019krf,Balitsky:2021cwr,Balitsky:2021qsr,Yao:2022vtp},
\textit{i.e.}  $\mathcal{O}(\alpha_s)$.  Furthermore, several efforts
have also been devoted to the calculation of higher-twist
contributions to off-light-cone
distributions~\cite{Radyushkin:2017ffo,Braun:2021gvv,Braun:2021aon,Braun:2023alc,Braun:2024snf,Han:2024cht}.
As a demonstration of the relevance of these quantities, a significant
number of studies have recently emerged which make use of these
kernels (see, \textit{e.g.},
Refs.~\cite{Alexandrou:2020zbe,Lin:2020rxa,Bhattacharya:2024qpp,HadStruc:2024rix} for recent lattice
QCD extractions of GPDs), along with first attempts to improve our
knowledge of hadron structure incorporating both experimental and
simulated data~\cite{Riberdy:2023awf,Karpie:2023nyg,Cichy:2024afd}.

Motivated by recent lattice extractions of heavy-meson DAs and
PDFs~\cite{Wang:2019msf,Zhao:2020bsx,Blossier:2024wyx,Blossier:2024eai,Han:2024fkr,LatticeParton:2024zko}, in this
paper we set out to incorporate heavy-quark mass effects into the
computation of the matching kernels. Specifically, our purpose is to
evaluate power corrections of $z^2m^2$ to the partonic quark
distributions with space-like separation $z^2$ in forward kinematics
(a.k.a. pseudo-distribution) of a heavy quark with mass $m$ up to
one-loop accuracy. This calculation will eventually allow us to
extract the matching kernels to connect the heavy-quark pseudo-distribution to PDFs.

The paper is organised as follows.  In section 2, the basic notation
is introduced. In section 3, the calculation of the one-loop quark-quark massive
pseudo-distribution is described. In section 4, the corresponding massive matching
kernel is extracted. In section 5, we present the contribution coming from the quark-gluon mixing. A numerical estimate of heavy-quark mass effects is presented in section 6. Finally, in section 7, we give a summary, draw our conclusions, and present an outlook. More details on the calculation are given in Appendices A and B.

\section{Ioffe-time distribution}
\label{Ioffet-time}

Let us start by considering the QCD quark string operator
\begin{equation}
   \mathcal{O}^{\alpha} = \bar{\psi} (0) \gamma^{\alpha} W(0,z,A) \psi (z) \; ,
   \label{Eq:Ioffe_OperatorDef}
\end{equation}
where
\begin{equation}
  W(0,z,A) = \mathcal{P}_{\rm exp} \left[ i g z_{\nu} \int_0^1 dt\,\hat{A}^{\nu} (tz) \right]
\end{equation}
is a straight-line gauge link in the fundamental representation. In
Ref.~\cite{Radyushkin:2017lvu}, it has been shown that the matching
kernel can be computed directly at the operator level in the
Balitsky-Braun spirit~\cite{Balitsky:1987bk}. However, since the
massive computation is particularly involved, we work at the level of
the distribution and compute the perturbative kernel using a target
quark. We thus consider the Ioffe-time distribution of a quark, which
reads
\begin{equation}
  \mathcal{M}^{\alpha} (\nu, z^2) = \frac{1}{2N_c} \sum_{c, \lambda} \langle p, \lambda | \bar{\psi} (0) \gamma^{\alpha} W(0,z,A) \psi (z) |p, \lambda \rangle \; ,   
\end{equation}
where the sum is over colour and quark helicities, and
$\nu = - p \cdot z$ is the Ioffe time, $p$ being the quark
momentum. The Ioffe-time distribution can be parametrised as
\begin{equation}
  \mathcal{M}^{\alpha} (\nu, z^2) = 2 p^{\alpha} f(\nu, z^2) + z^{\alpha} \tilde{f} (\nu, z^2) \; .
\end{equation}
We consider space-like separations, through the equal-time
parametrisation $z = (0,0,0,z^3)$, and the $\alpha=0$ component, which
allows us to avoid higher-twist contaminations in lattice
calculations.

At the leading order, the Wilson line is equal to the identity in
colour space and thus we have:
\begin{equation}
\begin{split}
    \mathcal{M}^{0} (\nu) & = \frac{1}{2 N_c} \sum_{c, \lambda} \langle p, \lambda | \bar{\psi} (0) \gamma^{0} \psi (z) |p, \lambda \rangle \\ & = 2 p^0 e^{i \nu} \equiv 2 p^{0} f (\nu) \; .
\end{split}
\end{equation}
The Fourier transform of $f (\nu)$ immediately gives the leading-order
quark parton distribution function (PDF)
\begin{equation}
    f (x) = \frac{1}{2 \pi} \int_{- \infty}^{+ \infty} d \nu \; e^{-i \nu x} f (\nu) = \delta (1-x) \; .
\end{equation}

In what follows, we extensively use the plus-prescription distribution
defined as
\begin{equation}
    \int_0^1 d \beta \; [f(\beta)]_+ \; g (\beta) = \int_0^1 d \beta  f(\beta) \left[ g (\beta) - g (1) \right] \; ,
\end{equation}
where $ f (\beta) \sim 1/(1-\beta)$ around $\beta = 1$, while
$g(\beta)$ is a regular function. Ultraviolet (UV) divergences are
always regularised in dimensional regularisation in
$D = 4 - 2 \epsilon_{ \rm UV}$ dimensions. Infrared (IR) divergences of the massive matrix element are instead absent due to the non-vanishing quark mass $m$. However, we point out that, in the self-energy of the massive fermion (section \ref{sec:Quarkline}) and in the box-like diagram (section \ref{sec:boxtype}), IR divergences associated with the gluon propagator appear which cancel out in the combination. For the sake of clarity, IR divergences will be regularised by making explicit $D=4-2 \epsilon_{\rm IR}$, while we keep $D$ implicit for UV divergences.

\section{One-loop calculation}

\subsection{Wilson-line self-energy contribution}
\label{sec:Wilson}

\begin{figure}
    \centering
    \includegraphics[width=0.5\linewidth]{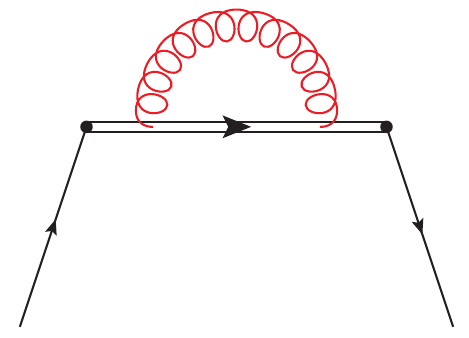}
    \caption{Wilson-line self-energy contribution.}
    \label{fig:WilsonLine}
\end{figure}

We work in Feynman gauge and thus we get a contribution from the
Wilson-line self-energy in fig.~\ref{fig:WilsonLine}. This
contribution vanishes on the light-cone, where $z^2 = 0$, and is
independent from the mass. It can be obtained directly at the operator
level by perturbatively expanding the Wilson line in eq.~(\ref{Eq:Ioffe_OperatorDef}) at the second order, \textit{i.e.}
\begin{align}
  \mathcal{O}^{\alpha}_{\rm Wils.} &= \bar{\psi} (0) \gamma^{\alpha} \frac{(ig)^2}{2} z^{\mu} z^{\nu} \nonumber \\
                                    & \times \int_0^1 d t_1 \int_0^1 d t_2 \left[ \theta (t_1 - t_2) (t^a t^b)_{i j} \wick{ \c2 A_{\mu}^a (t_1 z) \c2 A_{\nu}^b (t_2 z) } \right. \nonumber \\
                                    &\left. + \theta (t_2 - t_1) (t^b t^a)_{i j} \wick{ \c2 A_{\nu}^b (t_2 z) \c2 A_{\mu}^a (t_1 z)}  \right] \psi (z) . 
\end{align}
It is easy to show that
\begin{align}
  \mathcal{O}^{\alpha}_{\rm Wils.} &=  (i g)^2 \frac{C_F}{2} \int_0^1 d t_1 \int_0^1 d t_2 z^{\mu} z^{\nu} D_{\mu \nu} (z(t_1-t_2)) \nonumber \\ & \times \bar{\psi} (0) \gamma^{\alpha} \psi (z) 
   \equiv \Gamma_{\rm Wils.} (z) \bar{\psi} (0) \gamma^{\alpha} \psi (z) \; ,
\end{align}
where $D_{\mu \nu}$ is gluon propagator in position space, which in
dimensional regularisation reads
\begin{equation}
    D_{\mu \nu} (y) = - \frac{g_{\mu \nu} \Gamma (D/2-1)}{4 \pi^{D/2} (-y^2 + i0)^{D/2-1}} \; .
\end{equation}
Using the explicit form of the propagator, we immediately get
\begin{align}
  \Gamma_{\rm Wils.} (z) &= - \frac{g^2 C_F}{8 \pi^{D/2}} \Gamma \left(\frac{D}{2} -1 \right)  \nonumber \\
                           &  \times (-z^2) \int_0^1 d t_1 \int_0^1 d t_2 \frac{1}{\left[ -z^2 (t_1-t_2)^2 \right]^{D/2-1}} \; . 
\label{Eq:WilsonLineSelf_GammaW_1}
\end{align}
In $D=4$, the integral is divergent. There are several possible ways
to regularise it. In the seminal paper~\cite{Radyushkin:2017lvu},
these divergences were analysed using the Polyakov prescription
\begin{equation}
    \frac{1}{\left[ -z^2 (t_1-t_2)^2 \right]} \rightarrow \frac{1}{\left[ -z^2 (t_1-t_2)^2 + a^2 \right]} \; .
    \label{Eq:PolyakovPre}
\end{equation}
Besides standard logarithmic singularities, this prescription leads to
linear divergences that are interpreted as the renormalisation of a
mass moving along the gauge link. We do not enter into these
complications and rely on dimensional
regularisation to regularise the integral in eq.~(\ref{Eq:WilsonLineSelf_GammaW_1}). It is easy to see that
\begin{align}
  \Gamma_{\rm Wils.}  (z)  &= - \frac{g^2 C_F}{8 \pi^{D/2}} \Gamma \left(\frac{D}{2} -1 \right) (-z^2)^{2-D/2} \nonumber \\
  & \times \int_0^1 d t_1 \int_0^1 d t_2 \frac{1}{\left[ (t_1-t_2)^2 \right]^{D/2-1}} \; .
\end{align}
The integrals over $t_1$ and $t_2$ give
\begin{align}
  \int_0^1 \hspace{-0.1 cm} d t_1  \int_0^1 \hspace{-0.1 cm} d t_2 (t_1-t_2)^{2-D} =& \, 2 \int_0^1  \hspace{-0.1 cm} d t_1 \int_0^{t_1} \hspace{-0.1 cm} d t_2  (t_1-t_2)^{2-D} \nonumber \\
  = &\frac{2}{(D-3)(D-4)}
\end{align}
and we finally get
\begin{equation}
    \Gamma_{\rm Wils.} (z) = - \frac{g^2 C_F}{(4 \pi)^{D/2}} \frac{4 \; \Gamma \left(\frac{D}{2} -1 \right) }{(D-3)(D-4)} \left(\frac{-z^2}{4} \right)^{2-D/2} \; ,
\end{equation}
which agrees with the result of Refs.~\cite{Izubuchi:2018srq,Balitsky:2019krf}. For the one-loop contribution to the Ioffe-time distribution associated to the Wilson-line self-energy, we finally write
\begin{equation}
    \begin{split}
    \mathcal{M}_{\rm Wils.} (\nu, z^2) & = - \frac{g^2 C_F}{(4 \pi)^{D/2}} \frac{4 \; \Gamma \left(\frac{D}{2} -1 \right) }{(D-3)(D-4)} \left(\frac{-z^2}{4} \right)^{2-D/2}  \\ & \times  \mathcal{M}^{0} (\nu) \; .     
    \end{split}
    \label{Eq:Wilson_line_Contribution}
\end{equation}

\subsection{Quark-line self-energy contribution}
\label{sec:Quarkline}

\begin{figure}
    \centering
    \includegraphics[width=0.7\linewidth]{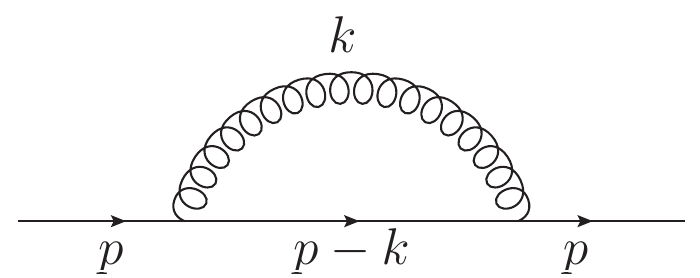}
    \caption{One-loop quark self energy.}
    \label{fig:Quark_Self_energy}
\end{figure}

To extract the quark self-energy contribution, we must consider the
one-loop quark propagator, which reads
\begin{align}
  &  i \left[D_{F} (p)\right]_{ij} =  \frac{i}{\slashed{p}-m} \delta_{ij} \nonumber \\
                                  & \times \left[ 1 +   \frac{i\left( -i \Sigma (p)  \right)}{\slashed{p}-m}   +   \frac{i\left[ i(Z_2 -1) \slashed{p} -i (Z_2 Z_m -1) m  \right]}{\slashed{p}-m} \right] , 
    \label{Eq:QuarkPropQCD}
\end{align}
where $Z_2$ and $Z_m$ are wave-function and mass renormalisation
constants, respectively. The factor $(-i \Sigma (p))$ is obtained from the amplitude
in fig.~\ref{fig:Quark_Self_energy} by amputating the external
spinors, \textit{i.e.}
\begin{align}
   - i \Sigma (p) \delta_{ij} & = \int \frac{d^D k}{(2 \pi)^D} (-i g \gamma^{\mu} t^a_{ik}) \frac{i \delta_{kn}}{\slashed{p}-\slashed{k}-m} 
 \nonumber \\ & \times (-i g \gamma^{\nu} t^b_{nj}) \frac{-i \delta^{ab}}{k^2} g_{\mu \nu} \;  .
\end{align}

After some algebra, one obtains
\begin{align}
  \Sigma (p) & = \frac{g^2 C_F }{(4 \pi)^{D/2}} 2 \Gamma \left( 2- \frac{D}{2} \right) \nonumber \\
               &\times \int_0^1 \textrm{d} \beta \left[ \frac{D}{2} m - \left(\frac{D}{2}-1 \right) (1-\beta) \slashed{p} \right] \nonumber \\ & \times ( \beta m^2 - \beta (1- \beta) p^2)^{\frac{D}{2} - 2}\,.
    \label{Eq:Self-Energy_FullSigma}
\end{align}
We have to include a correction for each external leg along with a
factor of 1/2 as a consequence of the LSZ reduction formula. Moreover,
in the presence of the double pole at $\slashed{p}=m$ in
eq.~(\ref{Eq:QuarkPropQCD}), the LSZ theorem prescribes that the one-loop correction to the Born diagram is \cite{Collins:2011zzd}
\begin{equation}
    \Gamma_{\rm Self.} = \frac{d \Sigma (p)}{d \slashed{p}} \bigg |_{\slashed{p}=m} \; .
  \end{equation}
In the case of a massive fermion, the self-energy correction features IR divergences, which may be regularised through an unphysical gluon mass. The gluon-mass regularisation is impractical in the calculation of the box diagram. For this reason, we adopt dimensional regularization. The derivative in eq.~(\ref{Eq:Self-Energy_FullSigma}) generates two terms: one is only UV-divergent, the other is only IR-divergent. We are thus entitled to use dimensional regularization with different regulators, $\epsilon_{\rm UV}$ and $\epsilon_{\rm IR}$, to extract the two singularities. We obtain
\begin{align}
    & \Gamma_{\rm Self.} = - \frac{2 g^2 C_F}{(4 \pi)^{D/2}} \left[ (m^2)^{D/2-2} \frac{\Gamma \left( 3 - \frac{D}{2} \right) }{\epsilon_{\rm IR}} + 2 \right]  \nonumber \\ & + \frac{g^2 C_F}{(4 \pi)^{D/2}} \left[ \frac{1}{\left( \frac{D}{2} - 2 \right)}  \left( \frac{-z^2}{4 e^{-\gamma_E}} \right)^{2-D/2} \hspace{-0.4 cm} + \ln \left( \frac{-z^2 m^2}{4 e^{-2 \gamma_E}} \right) \right]\nonumber \\ & + \mathcal{O} (D-4)\,.
   \label{Eq:MassiveQuarkSelfEnergy}
\end{align}
A first important observation is that the structure of the singularities of the massive self-energy is different from the massless case~\cite{Izubuchi:2018srq,Radyushkin:2017lvu}. Indeed, in the first line, we isolated an IR divergence which is absent in the massless computation. Moreover, as in the massless case \cite{Radyushkin:2017lvu}, we introduced a fictitious dependence on $z^2$.  \\

The self-energy contribution to the one-loop Ioffe-time distribution is finally written as
\begin{align}
   & \mathcal{M}_{\rm Self.} (\nu, z^2, m^2) \nonumber \\ & = - \frac{2 g^2 C_F}{(4 \pi)^{D/2}} \left[ (m^2)^{D/2-2} \frac{\Gamma \left( 3 - \frac{D}{2} \right) }{ \epsilon_{IR} } + 2 \right] \mathcal{M}^{0} (\nu)  \nonumber \\ 
                                            & + \frac{g^2 C_F}{(4 \pi)^{D/2}} \left[ \frac{1}{\left( \frac{D}{2} - 2 \right)}  \left( \frac{-z^2}{4 e^{-\gamma_E}} \right)^{2-D/2} \hspace{-0.4 cm} + \ln \left( \frac{-z^2 m^2}{4 e^{-2 \gamma_E}} \right) \right]\nonumber \\
                                            & \times \mathcal{M}^{0} (\nu)+ \mathcal{O} (D-4) \; .
   \label{Eq:Off-Light-Cone-Massive-Self}
\end{align}

\subsection{Box-type contribution}
\label{sec:boxtype}

\begin{figure}
  \centering
  \includegraphics[width=0.5\linewidth]{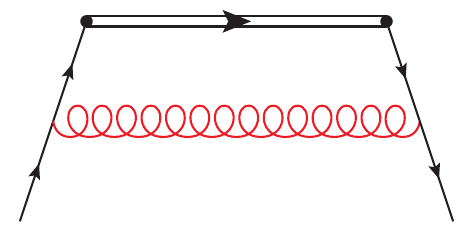}
  \caption{Box-like diagram contribution.}
  \label{fig:box}
\end{figure}

The correction to the operator coming from the box-like diagram in
fig.~\ref{fig:box} reads
\begin{align}
  & \mathcal{O}_{\rm Box}^{\alpha} = (ig)^2 C_F  \int d^D z_1 \int d^D z_2 \nonumber \\
                                   & \times \bar{\psi} (z_2) \gamma_{\mu} D_F (z_2) \gamma^{\alpha} D_F (z-z_1) \gamma_{\nu} \psi (z_1) D^{\mu \nu} (z_2-z_1) \; .
\end{align}
At the level of quark distribution, we have
\begin{align}
 & \mathcal{M}_{\rm Box} (\nu, z^2, m^2) \nonumber \\ & =   \frac{g^2 C_F \Gamma \left( \frac{D}{2} - 1 \right)}{8 \pi^{D/2}} \sum_{\lambda}  \int  d^D z_1 \hspace{-0.1 cm} \int \hspace{-0.1 cm} d^D z_2 e^{i p (z_2 -z_1)} \nonumber \\
  & \times  \frac{ \bar{u}_{\lambda} (p) \gamma_{\mu} D_F (z_2) \gamma^{0} D_F (z-z_1) \gamma^{\mu} u_{\lambda} (p) }{\left[ - (z_2-z_1)^2 + i0 \right]^{D/2 -1}} \; .
\end{align}
We then use the Fourier representation of the quark propagator
\begin{equation}
    D_F (z) =  \int \frac{d^D k}{(2 \pi)^D} e^{-i k z} \frac{i (\slashed{k}+m)}{k^2-m^2+i0} \; .
\end{equation}
After performing the shift $z_2 \rightarrow z_2 + z_1$ and integrating
over positions by means of the integral
\begin{equation}
   \int d^D z \frac{e^{i z (p-k_1)}}{\left[ - z^2/4 + i 0 \right]^{D/2-1}} = \frac{i (4 \pi)^{D/2}}{\Gamma \left( D/2 -1 \right)} \frac{1}{ \left[ (k_1 - p)^2 + i0 \right]} \; ,
   \label{Eq:Box_IntegralOverPos}
\end{equation}
we get
\begin{align}
     \mathcal{M}_{\rm Box} (\nu, z^2, m^2) 
  & =  \frac{g^2 C_F}{2} \int \frac{d^D k}{(2 \pi)^D i} e^{-i k z} \nonumber \\
  & \times \frac{ {\rm Tr} \left[ (\slashed{p} + m) \gamma_{\mu} (\slashed{k} + m) \gamma^0 (\slashed{k} + m) \gamma^{\mu} \right] }{\left[ (k-p)^2 + i0 \right] \left[ k^2 - m^2 + i0 \right]^2} \; .
    \label{Eq:BoxConStartPoint}
\end{align}
The calculation of the box contribution is lengthy. The general
strategy of the computation relies on using the Schwinger
representation for the denominators in eq.~(\ref{Eq:BoxConStartPoint})
to integrate over $k$. For compactness, we only provide the final
result and leave a detailed derivation to~\ref{App:BoxDet}. There we also check every possible limiting case of our result. \\

Off the light-cone, the box-like contribution does not have UV divergences, but, in analogy to the self-energy, it exhibits an IR divergence that can be traced back to the massless gluon dynamics. Therefore, we can set $D=4-2 \epsilon_{\rm IR}$ throughout and obtain
\begin{align}
  & \mathcal{M}_{\rm Box} (\nu, z^2, m^2) \nonumber \\
    & =  \frac{2g^2 C_F}{(4 \pi)^{D/2}} \left \{ \left[ (m^2)^{D/2-2} \Gamma \left( 3 - \frac{D}{2} \right) \left( \frac{1}{\epsilon_{\rm IR}} + 2 \right) \right. \right. \nonumber \\
  & \left. \left.
    +  2 \left( \frac{ 1 - \sqrt{-z^2 m^2} K_1 \left( \sqrt{-z^2 m^2 } \right) }{ -z^2 m^2  } \right) \right] \right. \mathcal{M}^{0} (\nu) \nonumber \\
    & +  \int_0^1 d \beta \left[ 2 (1-\beta) K_0 \left( \sqrt{-z^2 (1-\beta)^2 m^2 } \right) \right]_+ \mathcal{M}^{0} (\beta \nu)   \nonumber \\
                                          & + \frac{1}{2}  \int_0^1 d \beta \left[ \frac{-4 \beta}{1-\beta} \right]_+  \mathcal{M}^{0} ( \beta \nu) \nonumber \\ 
  &  \times \sqrt{-z^2 (1-\beta)^2 m^2 }  K_{1} \left( \sqrt{-z^2 (1-\beta)^2 m^2 } \right) \bigg \} \nonumber \\ & + \mathcal{O} (D-4) \; .
\label{Eq:BoxCompleteNoz2Exp2}
\end{align}
A few comments are in order. First, we observe that the
IR divergence in the second line of
eq.~(\ref{Eq:BoxCompleteNoz2Exp2}) cancels exactly that of the self-energy in eq.~(\ref{Eq:Off-Light-Cone-Massive-Self}), leaving an IR-finite result. The term in the third line of eq.~(\ref{Eq:BoxCompleteNoz2Exp2}) emerges
as a consequence of the fact that we enforced a plus-prescription
structure on the term proportional to
$K_0 ( \sqrt{-z^2 (1-\beta)^2 m^2 })$. In the massless calculation,
this term would cancel the logarithmic one in
eq.~(\ref{Eq:Off-Light-Cone-Massive-Self}). Indeed, the logarithmic
part of the two terms cancels exactly upon expansion around
$z^2m^2=0$, if one only retains the leading term (leading-power expansion). The third term in eq.~(\ref{Eq:BoxCompleteNoz2Exp2}) is
dominant in the limit $z^2 m^2 \rightarrow 0$ and, at leading power,
gives a contribution proportional to
\begin{equation}
\ln \left( \frac{4 e^{-2 \gamma_E} }{ -z^2 (1-\beta)^2 m^2 } \right) = \ln \left( \frac{4 e^{-2 \gamma_E}}{-z^2} \right) - \ln ((1-\beta)^2 m^2)\,.
\label{Eq:Log_Split}
\end{equation}
The first term on the r.h.s.~of eq.~(\ref{Eq:Log_Split}) is often referred to
as $z^2$-evolution term and is characteristic of pseudo-distributions. The second term contains the mass singularity and a finite term. We
observe that, besides the dominant logarithmic behaviour, accounting
for mass effects in the pseudo-distribution leads to a resummation of higher-power contributions proportional to powers of
$z^2 m^2$.\footnote{Each of these power terms also contains a
  $\ln(-z^2)$ according to the standard form of the operator
  product expansion.} Finally, the term proportional to \linebreak
$K_1 ( \sqrt{-z^2 (1-\beta)^2 m^2 })$ in the fifth and sixth lines of
eq.~(\ref{Eq:BoxCompleteNoz2Exp2}) is the combination of both
finite-mass and off-light-cone effects. At leading power, it has
no dependence on $z^2$, as expected. \\

We verified that, in the massless limit, our calculation reproduces
the expected result~\cite{Izubuchi:2018srq,Radyushkin:2017lvu}. It is important to note that this comparison
cannot be done naively starting from the massive result
and performing an expansion for $z^2 m^2 \rightarrow 0$. The correct
result is obtained by setting to zero all mass-related terms from the
start. The resulting integrals can then be computed in dimensional
regularisation with $D \neq 4$. We do not present here the comparison
diagram by diagram, but directly show the consistency between massive
and massless calculations for the full one-loop distribution.

\subsection{Vertex-type contribution}
\label{sec:vertextype}

\begin{figure}
    \centering
    \includegraphics[width=0.98\linewidth]{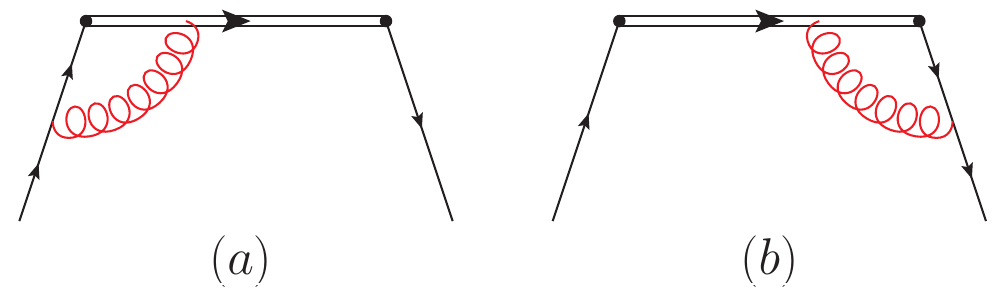}
    \caption{Vertex-like diagrams contribution.}
    \label{fig:VertexCon}
\end{figure}

The vertex correction is associated to the two diagrams in
fig.~\ref{fig:VertexCon}. The contribution of these two diagrams to
the forward matrix element is the same, therefore we consider only
$(b)$ and include $(a)$ by multiplying the result by a factor of
two. At the operator level, we have
\begin{align}
  \mathcal{O}_{\rm Vertex, b}^{\alpha} &= g^2 C_F \int_0^1 d t \int d^D z_1 \nonumber \\
  & \times D_{\mu \nu} ( z_1 - z t ) \bar{\psi} (z_1) \gamma^{\mu} D_F (z_1) \gamma^{\alpha}  \psi (z) z^{\nu} \; .
\end{align}
Moving to the quark distribution, we obtain
\begin{align}
  & \mathcal{M}_{\rm Vertex, b} (\nu, z^2, m^2) =  - \frac{g^2 C_F}{2} \frac{\Gamma \left( \frac{D}{2} -1 \right)}{(4 \pi)^{D/2}} i e^{i \nu} \\
                                               & \times \sum_{\lambda} \int_0^1 dt \int d^D z_1 \nonumber \int \frac{d^D k}{(2 \pi)^D} \frac{e^{i (p-k) z_1}}{\left[ - (z_1-zt)^2 / 4 + i0 \right]^{D/2 -1}} \nonumber \\
  & \times \bar{u}_{\lambda} (p) \slashed{z} \frac{\slashed{k} + m}{k^2-m^2 +i0} \gamma^{\alpha} u_{\lambda} (p) \; . 
\end{align}
Performing the shift $z_1 \rightarrow z_1 + z t$ and using the
integral in eq.~(\ref{Eq:Box_IntegralOverPos}), we get
\begin{align}
  & \mathcal{M}_{\rm Vertex, b} (\nu, z^2, m^2) =  \frac{g^2 C_F}{2} \int_0^1 d t \; e^{i \nu (1-t)}  \nonumber \\
                                                & \times \int \frac{d^D k}{(2 \pi)^D} \frac{e^{- i k \cdot z t} {\rm Tr} \left[ \slashed{z} \slashed{k} \gamma^{0} \slashed{p} \right]}{\left[ (k-p)^2 + i0 \right] \left[ k^2 - m^2 + i0 \right]} \; ,
    \label{Eq:VertexConFullyGen}
\end{align}
where we implicitly chose $z = (0,0,0,z^3)$. The vertex correction is the most complex, so we again defer a detailed derivation of the result to \ref{App:VertDet}. \\

The final result for the vertex contribution reads  
\begin{align}
  & \mathcal{M}_{\rm Vertex} (\nu, z^2, m^2) \nonumber \\
  & = \frac{g^2 C_F}{(4 \pi)^{D/2}} \bigg \{ -  \frac{2 \Gamma \left(\frac{D}{2}-1 \right) }{\left( \frac{D}{2} - 2 \right)}  \left( \frac{-z^2}{4} \right)^{2 - D/2} \hspace{-0.2 cm} \mathcal{M}^{0} (\nu)  \nonumber \\
  & + 2 \int_0^1 d \beta \left[ \frac{4 \beta}{ 1 - \beta } K_{0} \left( \sqrt{-z^2 (1-\beta)^2 m^2 } \right) \right]_+ \mathcal{M}^0 (\beta \nu)  \nonumber \\
  & - 2 \int_0^1 d \beta \left[ 4 \Phi ( 1 - \beta, \sqrt{-z^2 m^2} ) \mathcal{M}^{0} ( \beta \nu) \right.  \nonumber \\
  & \left. -4 \left( \frac{\ln ( 1 -\beta)+\beta}{1-\beta} \right) \; \mathcal{M}^{0} (\nu)  \right] - 8 \mathcal{M}^{0} (\nu) R \big( \sqrt{-z^2 m^2 } \big) \nonumber \\
  & +  \int_0^1 d \beta \left[ \frac{4 \beta}{1 - \beta} \right]_+  \mathcal{M}^0 (\beta \nu) \nonumber \\
  & \times \sqrt{-z^2 (1-\beta)^2 m^2} K_1 \left( \sqrt{-z^2 (1-\beta)^2 m^2} \right) \bigg \} \; ,
    \label{Eq:Vertex_Fin}
\end{align}
where 
\begin{align}
    & \Phi ( 1-\beta, \sqrt{-z^2 m^2} ) \equiv  \int_{1-\beta}^1 d t  \; \frac{\partial}{\partial \beta} \left[ \left( \frac{1-\beta}{t^2} - \frac{1}{t}  \right) \right. \nonumber \\
  & \times \left. \bigg ( K_0 \left( \sqrt{-z^2 m^2 (1-\beta)^2 }  \right) \right. \nonumber \\ & \left. \left. -  K_0 \left( \frac{ \sqrt{-z^2 m^2 (1-\beta)^2 } }{t} \right) \right) \right] \nonumber \\
  &  = \frac{\ln ( 1 -\beta)+\beta}{1-\beta} + \mathcal{O} (-z^2 m^2)\,,
\end{align}
\label{Eq:PhiFunc}
and 
\begin{align}
  & R \big( \sqrt{-z^2 m^2 } \big) =  \lim_{\beta \rightarrow 1} \int_{1-\beta}^1 \frac{d t}{t} \left( 1 - \frac{1-\beta}{t} \right) \Bigg[ \ln t  \nonumber \\ & + K_{0} \left( \sqrt{-z^2 m^2} (1-\beta) \right)  - K_{0} \left( \frac{ \sqrt{-z^2 m^2 } (1-\beta)}{t} \right) \Bigg] \; .
\end{align}
The UV-divergent term in the second line of eq.~(\ref{Eq:Vertex_Fin})
is identical to the massless case. The term proportional to
$K_{0} \left( \sqrt{-z^2 (1-\beta)^2 m^2 } \right)$ generalises the
$z^2$-evolution term of the massless case. Indeed, when combined with
an analogous term in the box-type correction, it produces the expected
structure
\begin{equation}
    \left[ \frac{1 + \beta^2}{ 1 - \beta} K_{0} \left( \sqrt{-z^2 (1-\beta)^2 m^2 } \right)  \right]_+\;.
\end{equation}
The term proportional to $K_1 ( \sqrt{-z^2 (1-\beta)^2 m^2 })$ exactly
cancels against the box and, in the leading-power approximation, it is
the \textit{UV-finite term} of Ref.~\cite{Radyushkin:2017lvu}. More
complications arise from the fourth and fifth line of
eq.~(\ref{Eq:Vertex_Fin}). A first important remark is that the
integrand in $\beta$ is finite when $\beta \rightarrow 1$. Indeed, the
function $\Phi ( 1-\beta, \sqrt{-z^2 m^2} ) $ is singular for
$\beta=1$ and its expansion around this value gives
\begin{align}
    \Phi ( 1-\beta, \sqrt{-z^2 m^2} )  = \frac{\beta + \ln ( 1 -\beta)}{1-\beta} + \mathcal{O} ((1-\beta)^0) \; .
\end{align}
Therefore, the whole square bracket involving the $\Phi$ function in
eq.~(\ref{Eq:Vertex_Fin}) is regular at $\beta=1$. Also, in the
leading-power approximation ($z^2 m^2 \rightarrow 0$), this term
reduces to the \textit{IR-finite term} of
Ref.~\cite{Radyushkin:2017lvu}. The function
$R \big( \sqrt{-z^2 m^2 } \big)$ in the fifth line of
eq.~(\ref{Eq:Vertex_Fin}) is finite and of $\mathcal{O} (-z^2 m^2)$
(thus absent in the massless limit). One may insist on computing $R$
in a closed form. However, it turned out to be easier to evaluate it
numerically. Moreover, in the pseudo-distribution approach, its exact
form is unimportant, since it cancels when taking the \textit{reduced}
Ioffe-time distribution~\cite{Radyushkin:2017lvu}. At the level of the RG-invariant ratio, even the term that removes the singular part of $\Phi$ vanishes, but this latter function becomes plus-prescribed, ensuring the finiteness of the result.
 
Finally, we checked the consistency with the massless limit~\cite{Izubuchi:2018srq,Radyushkin:2017lvu} also for
this contribution.

\subsection{Off-light-cone distribution at one-loop}

The complete massive Ioffe-time distribution is obtained by combining
eqs.~(\ref{Eq:Wilson_line_Contribution}),
(\ref{Eq:Off-Light-Cone-Massive-Self}),
(\ref{Eq:BoxCompleteNoz2Exp2}), and (\ref{Eq:Vertex_Fin}). We
find\footnote{Note that we added and subtracted a suitable
  contribution to isolate the higher-power part of the function $\Phi$
  in the last line of eq.~(\ref{Eq:FullResult}).}
\begin{align}
  & \mathcal{M}^{1-{\rm loop}} (\nu, z^2, m^2) = \frac{2 g^2 C_F}{(4 \pi)^{D/2}} \Bigg \{ Z (z^2) \mathcal{M}^{0} (\nu) \nonumber \\
  & +  \int_0^1  d \beta \left[ \frac{1 + \beta^2}{ 1 - \beta } 2 K_{0} \left( \sqrt{-z^2 (1-\beta)^2 m^2 } \right) \right. \nonumber \\ & \left. - 4 \frac{ \ln ( 1 -\beta) + \beta}{1-\beta} \right]_+ \hspace{-0.2 cm} \mathcal{M}^0 (\beta \nu) - 4 \hspace{-0.15 cm} \int_0^1 \hspace{-0.15 cm} d \beta \left[  \Phi ( 1 - \beta, \sqrt{-z^2 m^2} ) \right. \nonumber \\
  & \left. \left. -  \left( \frac{ \ln ( 1 -\beta) + \beta}{1-\beta} \right)  \right] \mathcal{M}^{0} ( \beta \nu) \right \} , 
    \label{Eq:FullResult}
\end{align}
where the function 
\begin{align}
 & Z (z^2)  =  -  \frac{(D-1)}{(D - 4) (D-3)}  \left( \frac{-z^2}{4 e^{-\gamma_E}} \right)^{2-D/2}  \nonumber \\
  & + 2 \left[ \frac{ 1 - \sqrt{-z^2 m^2} K_1 \left( \sqrt{-z^2 m^2 } \right) }{ -z^2  m^2  } \right. \nonumber \\ & \left. + \frac{1}{4} \ln \left( \frac{-z^2 m^2}{4 e^{-2 \gamma_E}} \right) -2 R \big( \sqrt{-z^2 m^2 } \big) \right]\,,
\end{align}
collects the (divergent) terms that drop when considering the reduced
Ioffe-time distribution~\cite{Radyushkin:2017lvu}.

\section{Quark-quark matching kernel}

Before building the massive quark-quark matching kernel, we inspect the leading
term in the $z^2 m^2 \rightarrow 0$ limit, to show the consistency
with the massless computation. We multiply eq.~(\ref{Eq:FullResult})
by a factor
\begin{equation}
   S_{\rm D} = \frac{(e^{\gamma_E} )^{2-D/2}}{(4 \pi)^{2-D/2}}
\end{equation}
to implement the $\overline{\rm MS}$ scheme and consider only the
leading term in the expansion around $z^2 m^2 = 0$, obtaining
\begin{align}
  & \mathcal{M}^{1-{\rm loop}} (\nu, z^2, m^2) \big |_{z^2 m^2 \rightarrow 0}\nonumber \\
  & = - \frac{ \bar{g}^2 C_F}{8 \pi^{2}} \Bigg \{ - \tilde{Z} (z^2) \mathcal{M}^{0} (\nu)    \nonumber \\
  &   + \hspace{-0.15 cm} \int_0^1 \hspace{-0.15 cm} d \beta \left[  \frac{ 4 \ln ( 1 -\beta) }{1-\beta} - 2 (1-\beta)  \right]_+ \mathcal{M}^{0} ( \beta \nu) \nonumber \\
  &  \hspace{-0.05 cm} + \int_0^1 \hspace{-0.15 cm} d \beta \left[ \frac{1 + \beta^2}{ 1 - \beta } \left( \ln \left( \frac{-z^2 m^2 }{4 e^{-2 \gamma_E-1}}  \right) + 2 \ln ( 1 - \beta ) + 1 \right) \right]_+ \nonumber \\ & \times \mathcal{M}^0 (\beta \nu) \Bigg \} \; , 
    \label{Eq:FullResultExp}
\end{align}
where $g = \bar{g} \mu^{\epsilon}$ and
\begin{equation}
   \tilde{Z} (z^2) = Z (z^2) \big |_{z^2 m^2 \rightarrow 0} = \frac{3}{2} \left( \frac{1}{\epsilon_{\rm UV}} + \ln \left( \frac{-z^2 \mu^2 }{4 e^{-2 \gamma_E}} \right) \right) + \frac{5}{2} \; .
\end{equation}
The massive pseudo-distribution, expanded for $z^2 m^2 \rightarrow 0$,
is almost identical to the massless one, which, adopting dimensional
regularisation also for the IR-sector,\footnote{For simplicity, we
  adopt a unique scale $\mu$.} reads~\cite{Izubuchi:2018srq,Radyushkin:2017lvu}
\begin{align}
  & \mathcal{M}^{1-{\rm loop}} (\nu, z^2, 0) \nonumber \\
   = &- \frac{ \bar{g}^2 C_F}{8 \pi^{2}} \Bigg \{  \hspace{-0.05 cm} - \tilde{Z} (z^2) \mathcal{M}^{0} (\nu) \nonumber \\ &  + \int_0^1 \hspace{-0.15 cm} d \beta \left[ \frac{1 + \beta^2}{ 1 - \beta } \left( \ln \left( \frac{-z^2 \mu^2 }{4 e^{-2 \gamma_E-1}}  \right) + \frac{1}{\epsilon_{\rm IR}} \right) \right]_+ \hspace{-0.15 cm} \mathcal{M}^0 (\beta \nu) \nonumber \\
  & + \hspace{-0.15 cm} \int_0^1 \hspace{-0.15 cm} d \beta \left[  \frac{ 4 \ln ( 1 -\beta) }{1-\beta} - 2 (1-\beta)  \right]_+ \mathcal{M}^{0} ( \beta \nu)  \Bigg \} \; . 
    \label{Eq:FullResultMassless}
\end{align} 
Both distributions in eqs.~(\ref{Eq:FullResultExp}) and
(\ref{Eq:FullResultMassless}) are now renormalised in the
$\overline{\rm MS}$ scheme by simply removing the UV pole in
$\tilde{Z} (z^2)$. It is clear that, in the massive case, the IR
pole is replaced by the logarithm of the mass and a finite term
proportional to $(2 \ln (1-\beta) +1)$ appeared.

This difference between the two Ioffe-time distributions is actually
correct. Indeed, while the massless Ioffe-time distribution must be
matched onto the massless light-cone Ioffe-time distribution
\begin{equation}
\begin{split}
  \mathcal{I}^{1-{\rm loop}} (\nu, \mu^2, 0) & = \frac{g^2}{8 \pi^2} C_F  \hspace{-0.15 cm}  \int_0^1 \hspace{-0.15 cm} d \beta  \left[ \frac{1+ \beta^2}{1 - \beta} \right]_+ \\ & \times \left( \frac{1}{\epsilon_{\rm UV}} - \frac{1}{\epsilon_{\rm IR}} \right) \mathcal{M}^{(0)} \left( \beta \nu \right) \; ,
\end{split}
 \label{Eq:Full_Massless}
\end{equation}
the massive version must be matched onto the massive ge\-ne\-ra\-li\-sa\-tion of
eq.~(\ref{Eq:Full_Massless}), which reads
\begin{align}
  & \mathcal{I}^{1-{\rm loop}} (\nu, \mu^2, m^2)  =  \frac{\bar{g}^2}{8 \pi^2} C_F  \int_0^1 \hspace{-0.15 cm} d \beta \; \mathcal{M}^0 (\beta \nu) \nonumber \\
  & \times \hspace{-0.05 cm} \left[  \frac{1+\beta^2}{1-\beta} \left( \frac{1}{\epsilon_{\rm UV}} - \ln \left( \frac{m^2}{\mu^2} \right) - 2  \ln (1-\beta) - 1 \right) \right]_+ .
\label{Eq:Light_Cone_Massive}
\end{align}
We observe that in eq.~(\ref{Eq:Light_Cone_Massive}), in addition to
the UV pole, we have the term
\begin{equation}
  \left[  \frac{1+\beta^2}{1-\beta} \left( - \ln \left( \frac{m^2}{\mu^2} \right) - 2  \ln (1-\beta) - 1 \right) \right]_+\; ,
    \label{Eq:MatchingVFNS}
\end{equation}
which is a known result in the context of the so-called heavy-quark
threshold matching relevant for PDF evolution in a
variable-flavour-number scheme~\cite{Ball:2015tna}.

As it can be seen by comparing eqs.~(\ref{Eq:FullResultExp})
and~(\ref{Eq:Light_Cone_Massive}) with
eqs.~(\ref{Eq:FullResultMassless}) and~(\ref{Eq:Full_Massless}), at
the level of the leading term in the $z^2 m^2 \rightarrow 0$ limit
(\textit{i.e.} the~leading-power approximation), the massive matching
kernel is the same as in the massless case. The difference is
therefore that the former accounts for higher-power corrections of the
type $z^2 m^2$ incorporated in the Bessel functions.

After these premises, we can build the complete matching of
eq.~(\ref{Eq:FullResult}) (consistently renormalised in the
$\overline{\rm MS}$ scheme) on eq.~(\ref{Eq:Light_Cone_Massive}) and
obtain
\begin{align}
  \label{eq:FinalKernel}
   & \mathcal{M}^{1-{\rm loop}} (\nu, z^2, m^2) \nonumber \\
 & =\mathcal{I}^{1-{\rm loop}} (\nu, \mu^2, m^2) + \frac{ \bar{g}^2 C_F}{8 \pi^{2}} \Bigg \{ Z_{\rm R} (z^2) \mathcal{I}^{0} (\nu)  \nonumber \\
   & + \hspace{-0.15 cm} \int_0^1 \hspace{-0.15 cm} d \beta \left[ \frac{1 + \beta^2}{ 1 - \beta } \left( 2 K_{0} \left( \sqrt{-z^2 (1-\beta)^2 m^2 } \right) + \ln \left( \frac{m^2}{\mu^2} \right) \right. \right. \nonumber \\
   & \left. \left.  + 2  \ln (1-\beta) + 1 \right) - 4 \frac{ \ln ( 1 -\beta) + \beta}{1-\beta} \right]_+ \hspace{-0.15 cm} \mathcal{I}^0 (\beta \nu) \nonumber \\
  & \left. - 4 \hspace{-0.15 cm} \int_0^1 \hspace{-0.15 cm} d \beta \left[  \Phi ( 1 - \beta, \sqrt{-z^2 m^2} )  -  \left( \frac{ \ln ( 1 -\beta) + \beta}{1-\beta} \right) \right] \right. \nonumber \\ &  \times \mathcal{I}^0 ( \beta \nu) \Bigg \} , 
\end{align}
where
\begin{align}
  Z_{\rm R} (z^2)  & = 2 + \frac{3}{2} \ln \left( \frac{-z^2 \mu^2}{4 e^{-2\gamma_E}} \right) \nonumber \\ & + 2 \left[ \frac{ 1 - \sqrt{-z^2 m^2} K_1 \left( \sqrt{-z^2 m^2 } \right) }{ -z^2  m^2  } \right.  \nonumber \\
  & \left.  + \frac{1}{4} \ln \left( \frac{-z^2 m^2}{4 e^{-2 \gamma_E}} \right) - 2 R \big( \sqrt{-z^2 m^2 } \big) \right] .  
\end{align}

\section{Gluon-quark matching kernel}
\label{sec:Gluon-quark mixing}

\begin{figure}
  \centering
  \includegraphics[width=0.5\linewidth]{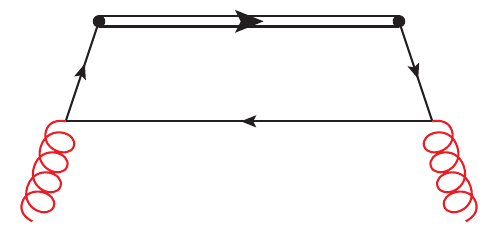}
  \caption{Diagram corresponding to the lowest-order contribution to the gluon-quark mixing.}
  \label{fig:boxGluon}
\end{figure}

The pseudo-distribution of the heavy quark mixes with that of the gluon. This contribution starts at one-loop order and the lowest-order diagram is given in fig.~\ref{fig:boxGluon}. The gauge-link structure, which differentiates the pseudo-distribution from the standard transverse-momentum-dependent distributions (TMDs), does not play any role in this diagram.\footnote{Pseudo-distributions and TMDs are defined through similar bi-local operators characterised by space-like separations and differ only by the gauge link. Specifically, pseudo-distributions feature a straight gauge link, while TMDs are defined through a staple-like gauge link in the light-cone direction~\cite{Collins:2011zzd}.} Therefore, this contribution coincides with that of heavy-quark TMDs~\cite{Nadolsky:2002jr,vonKuk:2023jfd}. In particular, we obtain
\begin{gather}
    \mathcal{M}_{\rm gluon-mix} (\nu, z^2, m^2) = \frac{\alpha_s}{2 \pi} 2 T_R \int_0^1 d \beta \; 2 p^0 f_g^{(0)} (\beta \nu) \nonumber \\ \times \left \{ [ \beta^2 + (1-\beta)^2 ] K_0 (\sqrt{-z^2 m^2}) \right. \nonumber \\ \left. + \beta (1-\beta) \sqrt{-z^2 m^2} K_1 (\sqrt{-z^2 m^2})  \right \} \; ,
\end{gather}
where $f_g^{(0)} (\nu) = e^{i \nu}$ is the leading-order Ioffe-time distribution of the gluon.

\section{Numerical analysis}

In order to quantitatively estimate the effect of heavy-quark mass
corrections on pseudo-distributions, we consider a proton target and
write the matching between heavy-quark pseudo-distribution and PDFs as follows:
\begin{gather}
 f_{Q} (x, z^2, \mu^2)  = \sum_{i=Q,g}\int_x^1 \frac{d y}{y} \mathcal{C}_{Qi} \left( y, z^2
   \mu^2, z^2 m^2, g \right) \nonumber \\ \times f_i \left( \frac{x}{y} , 0 , \mu^2 \right)
 \; ,
   \label{Eq:xspacematching}
\end{gather}
where, up to one-loop accuracy, we find
\begin{align}
  & \mathcal{C}_{QQ}\left( y, z^2 \mu^2, z^2 m^2, g \right) \nonumber \\
  & = \delta(1-y) + \frac{\bar{g}^2 C_F}{8 \pi^2} \Bigg \{ Z_{\rm R} (z^2 ) \delta (1-y) \nonumber \\ & + \left[ 
\frac{1+y^2}{1-y} \left( 2 K_0 \left( \sqrt{-z^2 m^2 (1-y)^2} \right)\right. \right.  \nonumber \\
  & + \left. \left. \ln \left( \frac{m^2}{\mu^2} \right) + 2 \ln (1-y) + 1 \right)
     - 4 \frac{\ln (1-y) + y}{1-y} \right]_+ \nonumber \\
  & - 4 \left( \Phi \left( 1-y, \sqrt{-z^2 m^2} \right) - \frac{\ln (1-y) +
    y}{1-y} \right) \Bigg \} \; ,
   \label{Eq:xspacekernel}
\end{align}
and
\begin{gather}
    \mathcal{C}_{Qg} \left( y, z^2 \mu^2, z^2 m^2, g \right) = \frac{\bar{g}}{8 \pi^2} 2 T_R  \\  \times \bigg \{ [ y^2 + (1-y)^2 ]   \left( K_0 (\sqrt{-z^2 m^2}) + \frac{1}{2} \ln \left( \frac{m^2}{\mu^2} \right) \right) \nonumber \\  + y (1-y) \sqrt{-z^2 m^2} K_1 (\sqrt{-z^2 m^2})  \bigg \} \; .
\end{gather}

\begin{figure}
    \centering
    \includegraphics[width=.99\linewidth]{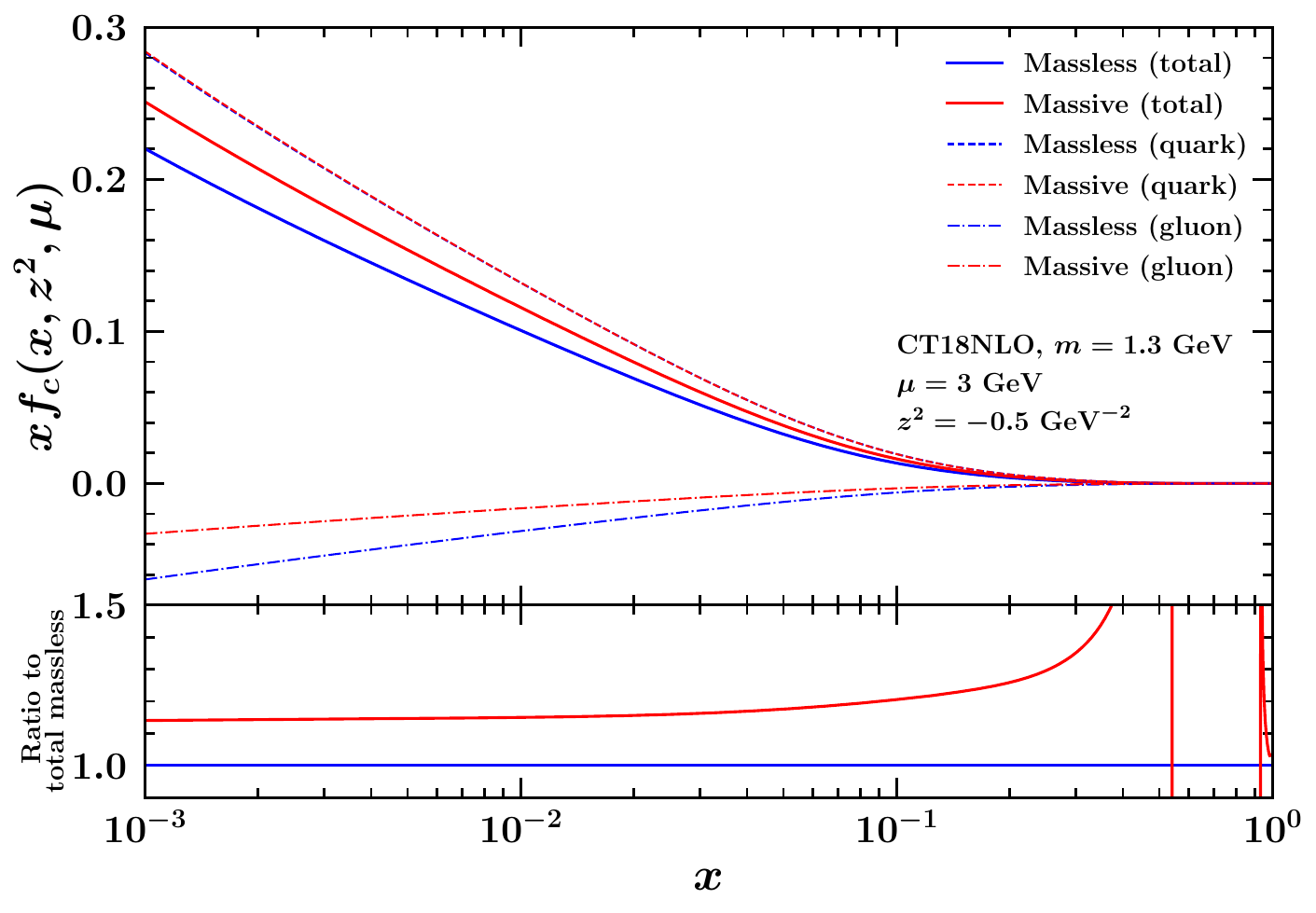}
    \caption{Charm-quark pseudo distribution as a function of the
      longitudinal momentum fraction $x$ computed by means of
      eq.~(\ref{Eq:xspacematching}) both in the massless approximation (blue curves) and with heavy-quark mass corrections (red curves).}
    \label{fig:CharmPseudoDistribution}
\end{figure}
In fig.~\ref{fig:CharmPseudoDistribution}, we present the charm
pseudo-distribution of the proton obtained by means of
eq.~(\ref{Eq:xspacematching}). In the computation, we used
PDFs and strong coupling from the {\tt CT18NLO}
set~\cite{Hou:2019efy}, accessed through the {\tt LHAPDF}
interface~\cite{Buckley:2014ana}. We set $\mu=3$~GeV, $m=1.3$~GeV, and
$z^2=-0.5$~GeV$^{-2}$. The upper panel of the plot shows the charm pseudo-distribution in the
massless limit (blue solid curve), and the massive charm
pseudo-distribution (red solid curve). We also show, for both massless and massive calculations, the separate contributions due to charm PDF (dashed curves) and gluon PDF (dot-dashed curves), corresponding to $i=Q$ and $i=g$ in eq.~(\ref{Eq:xspacematching}), respectively. The lower panel displays the total massive and massless curves normalised to the latter. By comparing massive and massless curves, we find that, at the level of total distributions, the inclusion of mass corrections generates an effect of approximately 15\% (see lower panel of fig.~\ref{fig:CharmPseudoDistribution}). The effect tends to grow as $x$ approaches one, where, however, the curves rapidly approach zero. We also find that the effect of heavy-quark mass
corrections is remarkably small on the quark-initiated channel, $i=Q$. Indeed, the dashed curves only differ by a few per mil across the full range in $x$ considered. We verified that a similar magnitude of differences is found for
other kinematic configurations. This observation indicates that there
is a strong and unexpected suppression of power corrections of the
form $z^2m^2$ in this channel. In contrast, mass effects on the gluon channel, $i=g$ (dot-dashed curves), are more sizeable. Specifically, we find that both massless and massive contributions produce negative results, with the former being twice as big in magnitude as the latter. Therefore, the gluon channel is almost entirely responsible for the difference between massless and massive pseudo-distributions.


\section{Summary and conclusions}
\label{sec:Summary}

We have computed the one-loop correction to the forward matrix element
of an off-light-cone bi-local quark correlator, often referred to as
pseudo-distribution, accounting for heavy-quark mass effects. This
calculation allowed us to extract the matching kernels on PDFs.

The computation of the quark-quark matching kernel is performed in Feynman gauge and features four
contributions: the Wilson-line self-energy (sec.~\ref{sec:Wilson}),
the quark-line self-energy (sec.~\ref{sec:Quarkline}), the box-type
contribution (sec.~\ref{sec:boxtype}), and the vertex-type
contribution (sec.~\ref{sec:vertextype}). The latter contribution
turned out to be the most challenging to compute. Indeed, the function
$\Phi$, introduced in eq.~(\ref{Eq:PhiFunc}) in a semi-analytical
fashion, is affected by a non-integrable end-point singularity that
needs to be treated with care. We proved that this singularity cancels
in the final result and made the cancellation apparent by expressing
the vertex-type contribution in a manifestly convergent form. The
final result, presented in eq.~\eqref{eq:FinalKernel}, resums all
powers of $z^2m^2$ by means of modified Bessel functions of the second
kind. We also noted that when
constructing the \textit{reduced} Ioffe-time distribution $Z_R$
cancels out. In addition, we presented the one-loop contribution to the gluon-to-heavy-quark distribution (sec.~\ref{sec:Gluon-quark mixing}), which is necessary to obtain the complete set on next-to-leading order corrections to the functions responsible for the matching of heavy-quark pseudo-distributions on PDFs.

In order to assess the quantitative impact of mass effects, we carried
out a numerical implementation of our result. After having extracted
the matching functions appropriate to construct the heavy-quark pseudo-distribution
in terms of PDFs, we evaluated the charm pseudo-distribution in the
proton using both our massive calculation and the known massless
calculation. When comparing the two results, we found that the impact
of quark-mass corrections amounts to around 15\% and is almost completely due to the quark-gluon mixing. This result suggests that this latter contribution must be included in the quantitative analysis of the unpolarized charm and bottom pseudo-distributions on the lattice. In the future, it will be interesting to explore the importance of these effects in the distributions of heavy
mesons~\cite{Zhao:2020bsx,Blossier:2024wyx,Blossier:2024eai} or in polarised pseudo-distributions.

\section*{Acknowledgements}

We are grateful to Benoît Blossier, Teseo San José, José Manuel
Morgado Chávez, Alessandro Papa, Simone Rodini, Samuel Wallon, Lech Szymanowski and Edoardo Spezzano for fruitful discussions. The work of V.B. is supported by l’Agence Nationale de la Recherche (ANR), project ANR-24-CE31-7061-01. The work of M.F. is
supported by the Agence Nationale de la Recherche under the contract
ANR-17-CE31-0019 and by the ULAM fellowship program of NAWA No. BNI/ULM/2024/1/00065 “Color glass condensate effective theory beyond the eikonal approximation”. M.F. also acknowledges support from the Italian
Foundation “Angelo Della Riccia”. The work of C.M. is supported, in
part, by l’Agence Nationale de la Recherche (ANR), project
ANR-23-CE31-0019.  This work was made possible by Institut Pascal at
Université Paris-Saclay with the support of the program
“Investissements d’avenir” ANR-11-IDEX-0003-01.  For the purpose of
open access, the authors have applied a CC-BY public copyright licence
to any Author Accepted Manuscript (AAM) version arising from this
submission. Feynman diagrams have been drawn using
JaxoDraw~\cite{Binosi:2008ig}.

\onecolumn

\appendix

\section{Box-diagram contribution (detailed derivation)}
\label{App:BoxDet}

In this appendix, we provide the on-light-cone limit of the box-diagram both in the massless and massive case, as well as a more detailed derivation of eq.~(\ref{Eq:BoxCompleteNoz2Exp2}). 

\subsection{On-light-cone limit}

We restart from eq.~(\ref{Eq:BoxConStartPoint}). To investigate the light-cone limit, it is useful to introduce the Sudakov basis
\begin{equation}
    n_{1}^{\mu} = \frac{1}{\sqrt{2}} \left( 1 , 0, 0, 1 \right) \; , \hspace{1 cm} n_{2}^{\mu} = \frac{1}{\sqrt{2}} \left( 1 , 0, 0, -1 \right)
\label{Eq:SudakovBasis}
\end{equation}
and parametrise the quark momenta as $p = p^+ n_1^{\mu} + \frac{m^2}{2 p^+} n_2^{\mu}$ and the distance between the quark field as $ z = z^{-} n_2^{\mu} $. After performing the change of variables $d^D k = d k^+ d k^{-} d^{D-2} \vec{k}_T$, eq.~(\ref{Eq:BoxConStartPoint}) becomes
\begin{gather}
     \mathcal{M}_{\rm Box} (\nu, 0, m^2) = \frac{g^2 C_F}{2} \int d k^+ e^{-i k^+ z^-} \int \frac{d k^{-}}{2 \pi i} \int \frac{d^{D-2} \vec{k}_T}{(2 \pi)^{D-1}}  \frac{1}{\left[ 2 k^+ k^- - \vec{k}_T^{\; 2} - m^2 + i0 \right]^2} \nonumber \\
     \times \frac{1}{\left[ 2 (k^+-p^+) (k^- - p^{-}) - \vec{k}_T^{\; 2} + i0 \right]} {\rm Tr} \left[ (\slashed{p} + m) \gamma_{\mu} (\slashed{k} + m) \gamma^0 (\slashed{k} + m) \gamma^{\mu} \right] \; .
    \label{Eq:BoxConAfterChange}
\end{gather}
\subsubsection*{Massless on-light-cone limit} 
We can perform the integral over $k^{-}$ using Cauchy's theorem. In the massless case, we get
\begin{gather}
      \int \frac{d k^{-}}{2 \pi i}  \frac{{\rm Tr} \left[ \slashed{p} \gamma_{\mu} \slashed{k} \gamma^0 \slashed{k}  \gamma^{\mu} \right]}{\left[ 2 (k^+-p^+) k^- - \vec{k}_T^{\; 2} + i0 \right] \left[ 2 k^+ k^- - \vec{k}_T^{\; 2} - m^2 + i0 \right]^2}  = \sqrt{2} (D-2) \left( 1 - \frac{k^+}{p^{+}} \right) \frac{1}{\vec{k}_T^{\; 2}} \theta (k^+) \theta (p^+ - k^+)  \; .
    \label{Eq:BoxConLightConeMassless}
\end{gather}
Then, eq.~(\ref{Eq:BoxConAfterChange}) reduces to
\begin{gather}
     \mathcal{M}_{\rm Box} (\nu, 0, 0)  = \frac{g^2 C_F}{2} (D-2) \sqrt{2} \int_0^{p^+} d k^+ e^{-i k^+ z^-} \left( 1 - \frac{k^+}{p^+} \right) \int \frac{d^{D-2} \vec{k}_T^{\; 2}}{(2 \pi)^{D-1}} \frac{1}{\vec{k}_T^{\; 2}} \; .
\end{gather}
The change of variables $\beta= k^+ / p^+$ leads us to
\begin{gather}
    \mathcal{M}_{\rm Box} (\nu, 0, 0)  = g^2 C_F \left( \frac{D}{2} - 1 \right) \int_0^1 (1-\beta) \mathcal{M}^{0} (\beta \nu) \int \frac{d^{D-2} \vec{k}_T^{\; 2}}{(2 \pi)^{D-1}} \frac{1}{\vec{k}_T^{\; 2}}\,.
\end{gather}
After the integral over the transverse momentum, introducing two different regulators for UV and IR divergences, we get 
\begin{equation}
    \mathcal{M}_{\rm Box} (\nu, 0, 0) = \frac{\alpha_s}{2 \pi} C_F \left( \frac{1}{\epsilon_{UV}} - \frac{1}{\epsilon_{IR}} \right) \int_0^1 d \beta \; (1 - \beta) \; \mathcal{M}^{(0)} \left( \beta \nu \right) \; .
    \label{Eq:BoxMasslessLightConeFin}
\end{equation}

\subsubsection*{Massive on-light-cone limit}

In the massive case, we can again use the Cauchy theorem to get
\begin{gather}
      \int \frac{d k^{-}}{2 \pi i}  \frac{ {\rm Tr} \left[ (\slashed{p} + m) \gamma_{\mu} (\slashed{k} + m) \gamma^0 (\slashed{k} + m) \gamma^{\mu} \right] }{\left[ 2 (k^+-p^+) (k^- - \frac{m^2}{2 p^+} ) - \vec{k}_T^{\; 2} + i0 \right] \left[ 2 k^+ k^- - \vec{k}_T^{\; 2} - m^2 + i0 \right]^2} \nonumber \\ = - \frac{1}{2 p^+}  \frac{\left( 1 - \frac{k^+}{p^+} \right)}{\left[ \vec{k}_T^{\; 2} + \left( 1 - \frac{k^+}{p^+} \right)^2 m^2 \right] } {\rm Tr} \left[ (\slashed{p} + m) \gamma_{\mu} (\slashed{k} + m) \gamma^0 (\slashed{k} + m) \gamma^{\mu} \right] \big |_{k^{-} = k_2^{-} } \; ,
    \label{Eq:BoxConLightConeMassive}
\end{gather}
where 
\begin{equation}
    k_2^{-}\equiv \frac{m^2}{2 p^+} - \frac{\vec{k}_T^{\; 2}}{2 (k^+ - p^+)} \; .
\end{equation}
The trace gives
\begin{gather}
    {\rm Tr} \left[ (\slashed{p} + m) \gamma_{\mu} (\slashed{k} + m) \gamma^0 (\slashed{k} + m) \gamma^{\mu} \right] \big |_{k^{-} = k_2^{-} } = - 2 \sqrt{2} p^+ \left[ (D-2) \left(\vec{k}_T^{\; 2} + \left(1- \frac{k^+}{p^+} \right)^2 m^2 \right) - 4 \frac{k^+}{p^+} m^2 \right] + \mathcal{O} \left( \frac{1}{p^+} \right)  .
\end{gather}
Thus, we obtain
\begin{gather}
      \int \frac{d k^{-}}{2 \pi i}  \frac{ {\rm Tr} \left[ (\slashed{p} + m) \gamma_{\mu} (\slashed{k} + m) \gamma^0 (\slashed{k} + m) \gamma^{\mu} \right] }{\left[ 2 (k^+-p^+) (k^- - \frac{m^2}{2 p^+} ) - \vec{k}_T^{\; 2} + i0 \right] \left[ 2 k^+ k^- - \vec{k}_T^{\; 2} - m^2 + i0 \right]^2} \nonumber \\ = \sqrt{2} \left( 1 - \frac{k^+}{p^+} \right) \left \{ \frac{(D-2)}{\left[ \vec{k}_T^{\; 2} + \left( 1 - \beta \right)^2 m^2 \right]}  - \frac{4 \beta m^2}{\left[ \vec{k}_T^{\; 2} + \left( 1 - \beta \right)^2 m^2 \right]^2} \right \} \; ,
\end{gather}
where we again introduced the variable $\beta = k^+ / p^+ $.
Therefore, eq.~(\ref{Eq:BoxConAfterChange}) becomes
\begin{gather}
\mathcal{M}_{\rm Box} (\nu, 0, m^2) = \frac{g^2 C_F}{2} \int_0^1 d \beta (1-\beta) \mathcal{M}^{0} ( \beta \nu)  \int \frac{d^{D-2} \vec{k}_T}{(2 \pi)^{D-1}} \left \{ \frac{(D-2)}{\left[ \vec{k}_T^{\; 2} + \left( 1 - \beta \right)^2 m^2 \right]}  - \frac{4 \beta m^2}{\left[ \vec{k}_T^{\; 2} + \left( 1 - \beta \right)^2 m^2 \right]^2} \right \} \; .
\end{gather}
The integrals over transverse momenta give
\begin{equation}
 \int \frac{d^{D-2} \vec{k}_T}{(2 \pi)^{D-1}} \frac{1}{\left[ \vec{k}_T^{\; 2} + \left( 1 - \beta \right)^2 m^2 \right]} = \frac{2}{(4 \pi)^{D/2}} \Gamma \left( 2 - \frac{D}{2} \right) (m^2)^{D/2-2} (1- \beta )^{D-4} 
 \label{Eq:FirstTransverseMomInt}
\end{equation}
and
\begin{equation}
 \int \frac{d^{D-2} \vec{k}_T}{(2 \pi)^{D-1}} \frac{1}{\left[ \vec{k}_T^{\; 2} + \left( 1 - \beta \right)^2 m^2 \right]^2} = \frac{2}{(4 \pi)^{D/2}} \Gamma \left( 3 - \frac{D}{2} \right) (m^2)^{D/2-3} (1- \beta )^{D-6} \; ,
\end{equation}
so that we finally get
\begin{gather}
\mathcal{M}_{\rm Box} (\nu, 0, m^2) = \frac{g^2 C_F}{ (4 \pi)^{D/2} } (m^2)^{D/2-2} \int_0^1 d \beta \mathcal{M}^{0} ( \beta \nu) \nonumber \\ \times \left[ \left( \Gamma \left(2 - \frac{D}{2} \right) (D-2) (1-\beta) - 4 \frac{\beta}{1-\beta} \Gamma \left(3 - \frac{D}{2} \right) \right) (1- \beta)^{D-4} \right] \; .
\label{Eq:BoxLightConeMass}
\end{gather}
In eq.~(\ref{Eq:BoxLightConeMass}), while the IR divergence associated with the quark dynamics has been regularised by the quark mass, an additional IR divergence associated with the gluon has appeared. Indeed, if $D=4$ is set, the second term is logarithmically divergent for $\beta \rightarrow 1$. This fact is not surprising; indeed, even the singularities of the self-energy are different from the massless case and, as we will see, these additional divergences cancel out. Using the relation
\begin{equation}
    \frac{4 \beta}{(1-\beta)^{5-D}} = \frac{4 \delta (1 - \beta)}{(D-4)} + \frac{4 \beta}{(1-\beta)_+} + \mathcal{O} (D-4)  \; ,
\end{equation}
after some algebra, we get
\begin{gather}
    \mathcal{M}_{\rm Box} (\nu, 0, m^2) = 2 \frac{g^2 C_F}{(4 \pi)^{D/2}} (m^2)^{D/2-2}   \Gamma \left( 2 - \frac{D}{2} \right)  \int_0^1 d \beta \bigg \{ \left[  (1-\beta)^{D-3}  \right]_+ + \left( \frac{D}{2} -2 \right) \left[ \frac{1+ \beta^2}{1-\beta} \right]_+ \bigg \} \mathcal{M}^0 (\beta \nu)  \nonumber \\ 
    + \frac{g^2 C_F}{(4 \pi)^{D/2}} (m^2)^{D/2-2} \bigg \{ \Gamma \left( 2 - \frac{D}{2} \right)  \mathcal{M}^{0} (\nu) + 2 \Gamma \left( 3 - \frac{D}{2} \right) \left( \frac{1}{\epsilon_{IR}} + 2 \right) \mathcal{M}^{0} (\nu) \bigg \} \; .
    \label{Eq:BoxLightConeMassiveFin}
\end{gather}

\subsection{Off-light-cone}
As explained in the body of the paper, it is easy to recover the massless limit from the final result, thus, we calculate the massive result directly.

\subsubsection*{Massive off-light-cone result}

We restart from eq.~(\ref{Eq:BoxConStartPoint}):
\begin{gather}
     \mathcal{M}_{\rm Box} (\nu, z^2, m^2) = \frac{g^2 C_F}{2} \int \frac{d^D k}{(2 \pi)^D i} e^{-i k z} \frac{ {\rm Tr} \left[ (\slashed{p} + m) \gamma_{\mu} (\slashed{k} + m) \gamma^0 (\slashed{k} + m) \gamma^{\mu} \right] }{\left[ (k-p)^2 + i0 \right] \left[ k^2 - m^2 + i0 \right]^2} \; .
\end{gather}
We use the Schwinger representation for the denominators
\begin{equation}
    \frac{1}{(A \pm i0)^n} = \frac{(\mp i)^n}{\Gamma (n)} \int_{0}^{\infty} d \sigma \sigma^{n-1} e^{\pm i \sigma (A \pm i0) } \; ,
    \label{Eq.SchwingerParam}
\end{equation}
to write
\begin{equation}
    \frac{1}{[(k-p)^2 + i0]} = - i \int_{0}^{\infty} d \sigma_2 e^{ i \sigma_2 ((k-p)^2 + i0 ) } \; , \hspace{0.2 cm} \frac{1}{ \left[ k^2 -m^2 + i0 \right]^2} = - \int_{0}^{\infty} d \sigma_1 \sigma_1 e^{ i \sigma_1 (k^2 -m^2 + i0) }
\end{equation}
and then perform the shift
\begin{equation}
    k \rightarrow k + \frac{z/2 + \sigma_2 p}{\sigma_1 + \sigma_2} \; ,
\end{equation}
to get
\begin{gather}
   \mathcal{M}_{\rm Box} (\nu, z^2, m^2) = i \frac{g^2 C_F}{2} \int_{0}^{\infty} d \sigma_1 \int_{0}^{\infty} d \sigma_2 \; \sigma_1 \; e^{-i \frac{\left( z/2 + \sigma_2 p \right)^2}{\sigma_1 + \sigma_2}} e^{-i (\sigma_1 - \sigma_2) m^2} \int \frac{d^D k}{(2 \pi)^D i} e^{i (\sigma_1 + \sigma_2) k^2 } \nonumber \\
   \times {\rm Tr} \left[ ( \slashed{p} + m) \gamma_{\mu} \left( \slashed{k} + \frac{ \frac{\slashed{z}}{2} + \sigma_2 \slashed{p}}{ (\sigma_1 + \sigma_2)} + m \right) \gamma^0 \left( \slashed{k} + \frac{ \frac{\slashed{z}}{2} + \sigma_2 \slashed{p}}{ (\sigma_1 + \sigma_2)}  + m \right) \gamma^{\mu} \right] \; .
  \label{Eq:BoxMassiveOffAfterShift}
\end{gather}
The Dirac trace in eq.~(\ref{Eq:BoxMassiveOffAfterShift}) has the form $A + B m^2$. Using the explicit parametrisation $z = (0,0,0,z^3)$ for the quark fields separation, the result can be written as the sum of three terms:
\begin{equation}
    \mathcal{M}_{\rm Box} (\nu, z^2, m^2) =  \mathcal{M}_{\rm Box, 1} (\nu, z^2, m^2) + \mathcal{M}_{\rm Box, 2} (\nu, z^2, m^2) + \mathcal{M}_{\rm Box, 3} (\nu, z^2, m^2) \; ,
\end{equation}
where
\begin{gather}
    \mathcal{M}_{\rm Box, 1} (\nu, z^2, m^2) = i \frac{g^2 C_F}{2} z^2 p^0 (D-2)  \int_0^{\infty} d \sigma_1 \int_0^{\infty} d \sigma_2 \frac{\sigma_1}{ (\sigma_1 + \sigma_2)^2} e^{-i \frac{\left( z/2 + \sigma_2 p \right)^2}{\sigma_1 + \sigma_2}} e^{-i (\sigma_1 - \sigma_2) m^2} \int \frac{d^D k}{(2 \pi)^D i} e^{i (\sigma_1 + \sigma_2) k^2 } \; ,
    \label{Mbox1Starting}
\end{gather}
\begin{gather}
    \mathcal{M}_{\rm Box, 2} (\nu, z^2, m^2) = i \frac{g^2 C_F}{2} (D-2) \int_0^{\infty} d \sigma_1 \int_0^{\infty} d \sigma_2 \; \sigma_1 \; \nonumber \\ \times  e^{-i \frac{\left( z/2 + \sigma_2 p \right)^2}{\sigma_1 + \sigma_2}} e^{-i (\sigma_1 - \sigma_2) m^2} \int \frac{d^D k}{(2 \pi)^D i} e^{i (\sigma_1 + \sigma_2) k^2 } \left[ 4 p^0 (k^2) - 4 k^0 (2 k \cdot p) \right] \; ,
     \label{Mbox2Starting}
\end{gather}
and 
\begin{gather}
    \mathcal{M}_{\rm Box, 3} (\nu, z^2, m^2) = - i g^2 C_F 2 p^0 m^2 \int_0^{\infty} d \sigma_1 \int_0^{\infty} d \sigma_2 \; \sigma_1 \; e^{-i \frac{\left( z/2 + \sigma_2 p \right)^2}{\sigma_1 + \sigma_2}} e^{-i (\sigma_1 - \sigma_2) m^2} \nonumber \\ \times   \left[ 2 \left(1-\frac{\sigma_2}{\sigma_1 + \sigma_2} \right)^2 - 4 \frac{\sigma_2}{\sigma_1 + \sigma_2} + (D-4) \left( 1-\frac{\sigma_2}{\sigma_1 + \sigma_2} \right)^2 \right] \int \frac{d^D k}{(2 \pi)^D i} e^{i (\sigma_1 + \sigma_2) k^2 } \; .
    \label{Mbox3Starting}
\end{gather}
We start by computing $\mathcal{M}_{\rm Box, 1} (\nu, z^2, m^2)$. We perform the $k$ integral, getting
\begin{equation}
    \int \frac{d^D k}{(2 \pi)^D i} e^{i (\sigma_1 + \sigma_2) k^2 } = \frac{(-i)^{D/2}}{(4 \pi)^{D/2}} \frac{1}{(\sigma_1 + \sigma_2)^{D/2}}
    \label{Eq:GaussianK}
\end{equation}
then we make the change of variables
\begin{equation}
    \beta = \frac{\sigma_2}{\sigma_1 + \sigma_2}  \hspace{0.5 cm} \lambda = \sigma_1 + \sigma_2 \; \implies \int_0^{\infty} \int_0^{\infty} d \sigma_1 d \sigma_2 \; ... = \int_0^{1} d \beta \int_0^{\infty} d \lambda \; \lambda \; ...
    \label{Eq:ConfTransf}
\end{equation}
In this way, we get\footnote{Note that, in presence of the heavy-quark mass, the $\mathcal{M}_{\rm Box, 1} (\nu, z^2, m^2)$ is free from divergences and thus we can set $D=4$.}
\begin{equation}
    \mathcal{M}_{\rm Box, 1} (\nu, z^2, m^2) = \frac{g^2 C_F}{2 (4 \pi)^{D/2}} \int_0^1 d \beta (1-\beta) \mathcal{M}^{0} (\beta \nu) (-i) z^2 \int_{0}^{\infty} \frac{d \lambda}{\lambda^2} \; e^{i \left(-\frac{ z^2}{4} \right) \frac{1}{\lambda} - i (1-\beta)^2 \lambda m^2} \; .
\end{equation}
Performing the inversion $\lambda \rightarrow 1 / \lambda$, the integral over $\lambda$ leads to a modified Bessel function,
\begin{equation}
    \int_{0}^{\infty} d \lambda \; e^{i \left(-\frac{ z^2}{4} \right) \lambda - i \frac{(1-\beta)^2}{\lambda} m^2} = 2 \left(  \frac{4 m^2}{z^2} \right)^{1/2} K_1 \left( \sqrt{-z^2 (1-\beta)^2 m^2 } \right) \; ,
    \label{Eq:MacDonald1}
\end{equation}
and we finally obtain
\begin{equation}
    \mathcal{M}_{\rm Box, 1} (\nu, z^2, m^2) = - \frac{2 g^2 C_F}{(4 \pi)^{D/2}} \int_0^1 d \beta (1-\beta) \mathcal{M}^{0} (\beta \nu) \sqrt{-z^2 (1-\beta)^2 m^2 } K_1 \left( \sqrt{-z^2 (1-\beta)^2 m^2 } \right) .
    \label{Eq:MBox1}
\end{equation}
We now move to $\mathcal{M}_{\rm Box, 2} (\nu, z^2, m^2)$. In this case, we perform the integral over the 4D Minkowskian components by using generalised Gaussian integrals, \textit{i.e.}
\begin{gather}
    \int \frac{d k^0 d k^1 d k^2 d k^3}{(2 \pi)^D i} e^{i (\sigma_1 + \sigma_2) \left[ (k^0)^2 - (k^1)^2 - (k^2)^2 - (k^3)^2 \right]} \left[  (k^0)^2 + (k^1)^2 + (k^2)^2 + (k^3)^2 \right] = \frac{i}{(4 \pi)^2} \frac{1}{(\sigma_1 + \sigma_2)^3} \; . 
\end{gather}
Then, by using the change of variables in (\ref{Eq:ConfTransf}) and the transformation $\lambda \rightarrow 1 / \lambda$, we obtain
\begin{equation}
   \mathcal{M}_{\rm Box, 2} (\nu, z^2, m^2) = \frac{2 g^2 C_F}{( 4 \pi)^{D/2}} \int_0^1 d \beta (1-\beta) \mathcal{M}^{0} (\beta \nu) \int_{0}^{\infty} \frac{d \lambda}{\lambda} \; e^{i \left(-\frac{ z^2}{4} \right) \lambda - i \frac{(1-\beta)^2}{\lambda} m^2}\,.
\end{equation}
The integral over $\lambda$ again gives a modified Bessel function, \textit{i.e.}
\begin{equation}
    \int_{0}^{\infty} \frac{d \lambda}{\lambda} \; e^{i \left(-\frac{ z^2}{4} \right) \lambda - i \frac{(1-\beta)^2}{\lambda} m^2} = 2 K_0 \left( \sqrt{-z^2 (1-\beta)^2 m^2 } \right) \; ,
    \label{Eq:MacDonald2}
\end{equation}
so that we finally get
\begin{equation}
   \mathcal{M}_{\rm Box, 2} (\nu, z^2, m^2) = \frac{2 g^2 C_F}{( 4 \pi)^{D/2}} \int_0^1 d \beta (1-\beta) \mathcal{M}^{0} (\beta \nu)  2 K_0 \left( \sqrt{-z^2 (1-\beta)^2 m^2 } \right) \; .
   \label{Eq:MBox2}
\end{equation}
We finally consider $\mathcal{M}_{\rm Box, 3} (\nu, z^2, m^2)$ in eq.~(\ref{Mbox3Starting}). In this case, it is important to keep $D \neq 4$. Using in sequence (\ref{Eq:GaussianK}), (\ref{Eq:ConfTransf}), and the inversion $\lambda \rightarrow 1 / \lambda$, we get
\begin{gather}
    \mathcal{M}_{\rm Box, 3} (\nu, z^2, m^2) = (- i)^{1+D/2} \frac{g^2 C_F}{(4 \pi)^{D/2}} m^2 \int_0^{1} d \beta (1-\beta) \mathcal{M}^{0} (\beta \nu) \nonumber \\ \times \int_0^{\infty} d \lambda \; \lambda^{D/2 -4} e^{i \left( \frac{-z^2}{4} \right) \lambda - i \frac{(1-\beta)^2}{\lambda} m^2 }  \left[ 2 \left( 1 - \beta \right)^2 + (D-4) \left( 1 - \beta \right)^2 - 4 \beta \right] \; .
\end{gather}
Since only the last term in the square brackets has a singularity (when $\beta \rightarrow 1$) and the others are finite, we can neglect the term proportional to $(D-4)$. Performing the integral over $\lambda$ through the formula
\begin{equation}
    \int_0^{\infty} d \lambda \; \lambda^{D/2 -4} e^{i \left( \frac{-z^2}{4} \right) \lambda - i \frac{(1-\beta)^2}{\lambda} m^2 } = 2 \left( \frac{4 m^2 (1-\beta)^2}{z^2} \right)^{D/4 - 3/2} K_{D/2 -3} \left( \sqrt{-z^2 (1-\beta)^2 m^2 } \right) \; ,
\end{equation}
we obtain
\begin{gather}
    \mathcal{M}_{\rm Box, 3} (\nu, z^2, m^2) = \frac{2 g^2 C_F}{(4 \pi)^{D/2}} \int_0^1 d \beta (1-\beta) \mathcal{M}^{0} (\beta \nu) \sqrt{-z^2 (1-\beta)^2 m^2 } K_1 \left( \sqrt{-z^2 (1-\beta)^2 m^2 } \right) \nonumber \\ + \frac{g^2 C_F}{(4 \pi)^{D/2}} (m^2)^{D/2-2} \int_0^1 d \beta \left[ \frac{-4 \beta}{(1-\beta)^{5-D}} \right] \mathcal{M}^{0} ( \beta \nu)  2^{D/2-2} \left[ \sqrt{-z^2 (1-\beta)^2 m^2 } \right]^{3-D/2} K_{D/2-3} \left( \sqrt{-z^2 (1-\beta)^2 m^2 } \right)  ,
    \label{Eq:MBox3Fin}
\end{gather}
where, in the non-divergent term, we set $D=4$. Summing all $\mathcal{M}_{\rm Box, i} (\nu, z^2)$, we obtain the following result:
\begin{gather}
    \mathcal{M}_{\rm Box} (\nu, z^2, m^2) = \frac{2 g^2 C_F}{( 4 \pi)^{D/2}} \int_0^1 d \beta (1-\beta) \mathcal{M}^{0} (\beta \nu)  2 K_0 \left( \sqrt{-z^2 (1-\beta)^2 m^2 } \right) \nonumber \\ + \frac{g^2 C_F}{(4 \pi)^{D/2}} (m^2)^{D/2-2} \int_0^1 d \beta \left[ \frac{-4 \beta}{(1-\beta)^{5-D}} \right] \mathcal{M}^{0} ( \beta \nu) 2^{D/2-2} \left[ \sqrt{-z^2 (1-\beta)^2 m^2 } \right]^{3-D/2} K_{D/2-3} \left( \sqrt{-z^2 (1-\beta)^2 m^2 } \right)  .
    \label{Eq:MBoxMassz2BeforePlusMan}
\end{gather}
Since 
\begin{equation}
  \lim_{\beta \rightarrow 1} 2^{D/2-2} \left[ \sqrt{-z^2 (1-\beta)^2 m^2 } \right]^{3-D/2} K_{D/2-3} \left( \sqrt{-z^2 (1-\beta)^2 m^2 } \right) = \Gamma \left( 3 - \frac{D}{2} \right) \; ,
\end{equation}
we treat it as a regular test function and use the relation
\begin{gather}
    \frac{-4 \beta}{(1-\beta)^{5-D}} = - 4 \frac{\delta (1-\beta) }{D-4} - \frac{4 \beta}{(1-\beta)_+} + \mathcal{O} (D-4) 
    = - 4 \delta ( 1 - \beta ) \left( \frac{1}{D-4} - 1 \right) - \left[ \frac{4 \beta}{1-\beta} \right]_{+} + \mathcal{O} (D-4) \; ,
\end{gather}
valid in the distributional sense, to obtain
\begin{gather}
    \mathcal{M}_{\rm Box} (\nu, z^2, m^2) = \frac{g^2 C_F}{(4 \pi)^{D/2}} (m^2)^{D/2-2} \Gamma \left( 3 - \frac{D}{2} \right) \mathcal{M}^{0} (\nu) \left( \frac{-4}{D-4} + 4 \right) \nonumber \\
    + \frac{2 g^2 C_F}{( 4 \pi)^{D/2}} \int_0^1 d \beta (1-\beta) \mathcal{M}^{0} (\beta \nu)  2 K_0 \left( \sqrt{-z^2 (1-\beta)^2 m^2 } \right) \nonumber \\
    + \frac{g^2 C_F}{(4 \pi)^{D/2}}  \int_0^1 d \beta \left[ \frac{-4 \beta}{1-\beta} \right]_+  \mathcal{M}^{0} ( \beta \nu)  \sqrt{-z^2 (1-\beta)^2 m^2 }  K_{1} \left( \sqrt{-z^2 (1-\beta)^2 m^2 } \right) \; .
    \label{Eq:BoxCompleteNoz2Exp}
\end{gather}
We can force the appearance of the plus prescription in the second line of eq.~(\ref{Eq:BoxCompleteNoz2Exp}) and obtain 
\begin{gather}
    \mathcal{M}_{\rm Box} (\nu, z^2, m^2) = \frac{g^2 C_F}{(4 \pi)^{D/2}} (m^2)^{D/2-2} \Gamma \left( 3 - \frac{D}{2} \right) \mathcal{M}^{0} (\nu) \left( \frac{-4}{D-4} + 4 \right) \nonumber \\
    + \frac{4 g^2 C_F}{( 4 \pi)^{D/2}} \left[ \frac{ 1 - \sqrt{-z^2 m^2} K_1 \left( \sqrt{-z^2 m^2 } \right) }{ -z^2 m^2  } \right] \mathcal{M}^{0} (\nu) 
    + \frac{2 g^2 C_F}{( 4 \pi)^{D/2}} \int_0^1 d \beta \left[ 2 (1-\beta) K_0 \left( \sqrt{-z^2 (1-\beta)^2 m^2 } \right) \right]_+ \mathcal{M}^{0} (\beta \nu)   \nonumber \\
    + \frac{g^2 C_F}{(4 \pi)^{D/2}}  \int_0^1 d \beta \left[ \frac{-4 \beta}{1-\beta} \right]_+  \mathcal{M}^{0} ( \beta \nu)  \sqrt{-z^2 (1-\beta)^2 m^2 }  K_{1} \left( \sqrt{-z^2 (1-\beta)^2 m^2 } \right) \; ,
\label{Eq:BoxCompleteNoz2Exp2_Ape}
\end{gather}
which coincides with eq.~(\ref{Eq:BoxCompleteNoz2Exp2}).

\section{Vertex-diagram contribution (detailed derivation)}
\label{App:VertDet}

In this appendix, we provide the on-light-cone limit of the vertex-diagram both in the massless and massive case, as well as a more detailed derivation of eq.~(\ref{Eq:Vertex_Fin}). 

\subsection{On-light-cone limit}

To investigate the light-cone distribution, we introduce again the Sudakov basis (\ref{Eq:SudakovBasis}) and perform the change of variables $d^D k = d k^+ d k^{-} d^{D-2} \vec{k}_T$.

\subsubsection*{Massless on-light-cone limit}

In the massless limit, after integrating over $k^-$ and performing the change of variables $\beta = k^+ / p^+ $, we get
\begin{gather}
    \mathcal{M}_{\rm Vertex, b} (\nu, 0, 0) = g^2 C_F \int_0^1 d \beta \frac{\beta}{1-\beta} 2 p^0 e^{i \nu} \int_0^1 dt \; (-i \nu) (1-\beta) e^{- i \nu ( 1 - \beta) t } \int \frac{d^{D-2} \vec{k}_T}{(2 \pi)^{D-1}} \frac{1}{\vec{k}_T^{\; 2}} \nonumber \\ = g^2 C_F \int_0^1 d \beta \frac{\beta}{1-\beta} \left( \mathcal{M}^0 (\beta \nu) - \mathcal{M}^0 ( \nu) \right) \int \frac{d^{D-2} \vec{k}_T}{(2 \pi)^{D-1}} \frac{1}{\vec{k}_T^{\; 2}}\,.
    \label{Eq:Vertex_Massless_Inter}
\end{gather}
Performing the integral over the transverse momentum and including the contribution from the diagram $(a)$ in fig.~\ref{fig:VertexCon}, we get
\begin{equation}
    \mathcal{M}_{\rm Vertex} (\nu, 0, 0) = \frac{\alpha_s}{2 \pi} C_F \left( \frac{1}{\epsilon_{UV}} - \frac{1}{\epsilon_{IR}} \right) \int_0^1 d \beta \; \left[ \frac{2 \beta}{1-\beta} \right]_+ \; \mathcal{M}^{(0)} \left( \beta \nu \right) \; .
    \label{Eq:Vertex_LightCone_Massless_Fin}
\end{equation}
Combining eqs.~(\ref{Eq:BoxMasslessLightConeFin}) and (\ref{Eq:Vertex_LightCone_Massless_Fin}) and adding the massless self-energy diagram calculated in full dimensional regularisation, we obtain the standard result for the one-loop massless PDF of eq.~(\ref{Eq:Full_Massless}). 

\subsubsection*{Massive on-light-cone limit}

To obtain the massive result it is enough to perform the replacement
\begin{equation}
    \int \frac{d^{D-2} \vec{k}_T}{(2 \pi)^{D-1}} \frac{1}{\vec{k}_T^{\; 2}} \longrightarrow \int \frac{d^{D-2} \vec{k}_T}{(2 \pi)^{D-1}} \frac{1}{ \left[ \vec{k}_T^{\; 2} + (1-\beta)^2 m^2 \right]} \; ,
\end{equation}
in the first line of~(\ref{Eq:Vertex_Massless_Inter}). Then, using the integral in eq.~(\ref{Eq:FirstTransverseMomInt}), we have
\begin{equation}
    \mathcal{M}_{\rm Vertex} (\nu, 0, m^2)  = \frac{g^2 C_F}{ (4 \pi)^{D/2} } \Gamma \left( 2 - \frac{D}{2} \right) (m^2)^{D/2-2} \int_{0}^{1} d \beta \left[ \frac{4 \beta}{1-\beta} (1-\beta)^{D-4} \right]_+ \; \mathcal{M}^{(0)} \left( \beta \nu \right) .
    \label{Eq:Vertex_Massive_LightConeFin}
\end{equation}
Combining eqs.~(\ref{Eq:BoxLightConeMassiveFin}), (\ref{Eq:Vertex_Massive_LightConeFin}), and adding the massive self-energy in eq.~(\ref{Eq:MassiveQuarkSelfEnergy}), we recover the one-loop massive PDF of eq.~(\ref{Eq:Light_Cone_Massive}).

\subsection{Off-light-cone}

Again, it is easy to recover the massless limit from the final result and we thus calculate the massive result directly.

\subsubsection*{Massive off-light-cone result}
We restart form eq.~(\ref{Eq:VertexConFullyGen}),
\begin{gather}
    \mathcal{M}_{\rm Vertex, b} (\nu, z^2, m^2) = \frac{g^2 C_F}{2} \int_0^1 d t \; e^{i \nu (1-t)} \int \frac{d^D k}{(2 \pi)^D} \frac{e^{- i k \cdot z t} {\rm Tr} \left[ \slashed{z} \slashed{k} \gamma^{0} \slashed{p} \right]}{\left[ (k-p)^2 + i0 \right] \left[ k^2 - m^2 + i0 \right]} \; .
\end{gather}
Using the Schwinger parametrization (\ref{Eq.SchwingerParam}) for the denominators and the integral in eq.~(\ref{Eq:GaussianK}), we obtain
\begin{gather}
    \mathcal{M}_{\rm Vertex, b} (\nu, z^2, m^2) = \frac{g^2 C_F}{(4 \pi)^{D/2}} (-i)^{1+D/2} \; p^0 \int_0^1 dt \; e^{i (1-t) \nu} \int_0^{\infty} d \sigma_1 \int_0^{\infty} d \sigma_2 \frac{1}{(\sigma_1 + \sigma_2)^{D/2}} \nonumber \\ \times e^{-i \frac{\sigma_2^2}{\sigma_1 + \sigma_2} m^2 + i \frac{1}{ (\sigma_1 + \sigma_2)} \left( \frac{-z^2  t^2}{4} \right) + i \frac{\sigma_1}{\sigma_1 + \sigma_2} \nu \; t } \left( \frac{z^2 t}{\sigma_1 + \sigma_2} - 4 \frac{\sigma_1}{\sigma_1 + \sigma_2} \nu \right)  \; .
\end{gather}
The change of variables in eq.~(\ref{Eq:ConfTransf}) allow us to cast the result as
\begin{gather}
    \mathcal{M}_{\rm Vertex, b} (\nu, z^2, m^2) = \frac{g^2 C_F}{(4 \pi)^{D/2}}  (-i)^{1+D/2} \; p^0 \int_0^1 dt \; e^{i (1-t) \nu} \int_0^1 d \beta \int_{0}^{\infty} d \lambda \; \lambda^{1-D/2} \nonumber \\ \times e^{-i (1-\beta)^2 m^2 \lambda + i \left( \frac{-z^2  t^2}{4} \right) \frac{1}{\lambda} + i \beta \nu \; t } \left( \frac{z^2 t}{\lambda} - 4 \beta \nu \right) \; .
    \label{Eq:Vertex_Off-Light-Cone_Inter}
\end{gather}
After performing the inversion $\lambda \rightarrow 1/\lambda$, the integral over $\lambda$ can be taken using the known integrals
\begin{equation}
    \int_0^{\infty} d \lambda \; \lambda^{D/2 - 2} e^{i \left( \frac{-z^2 t^2}{4} \right) \lambda - i \frac{(1-\beta)^2}{\lambda} m^2 } = 2 \left( \frac{4 m^2 (1-\beta)^2}{z^2 t^2} \right)^{\frac{D-2}{4}} K_{\frac{D-2}{2}} \left( \sqrt{-z^2 t^2 (1-\beta)^2 m^2 } \right) 
\end{equation}
and
\begin{equation}
    \int_0^{\infty} d \lambda \; \lambda^{D/2 - 3} e^{i \left( \frac{-z^2 t^2}{4} \right) \lambda - i \frac{(1-\beta)^2}{\lambda} m^2 } = 2 \left( \frac{4 m^2 (1-\beta)^2}{z^2 t^2} \right)^{\frac{D-4}{4}} K_{\frac{D-4}{2}} \left( \sqrt{-z^2 t^2 (1-\beta)^2 m^2 } \right) \; \; .
\end{equation}
Then, we obtain
\begin{gather}
    \mathcal{M}_{\rm Vertex, b} (\nu, z^2, m^2) = \frac{g^2 C_F}{(4 \pi)^{D/2}}  (-i)^{1+D/2} \; p^0 \int_0^1 dt \; \int_0^1 d \beta e^{i (1-t) \nu + i \beta t \nu} 2 \left( \frac{4 m^2 (1-\beta)^2}{z^2 t^2}  \right)^{\frac{D-4}{4}} \nonumber \\ \times \left[ z^2 t \left( \frac{4 m^2 (1-\beta)^2}{z^2 t^2} \right)^{1/2} K_{\frac{D-2}{2}} \left( \sqrt{-z^2 t^2 (1-\beta)^2 m^2 } \right) - 4 \beta \nu K_{\frac{D-4}{2}} \left( \sqrt{-z^2 t^2 (1-\beta)^2 m^2 } \right) \right] \; .
    \label{OffLightConeVertex}
\end{gather}
We calculate separately the two terms in the square bracket of eq.~(\ref{OffLightConeVertex}). The first one clearly contains a singularity when $t \rightarrow 0$. This is an UV divergence and, in order to isolate it, we add and subtract a suitable term. The UV-singular part is obtain by taking the $t \rightarrow 0$ limit of the integrand in the first term in eq.~(\ref{OffLightConeVertex}), \textit{i.e.}
\begin{gather}
    \mathcal{M}_{\rm Vertex, b, UV-sing.} (\nu, z^2, m^2) = \frac{g^2 C_F}{(4 \pi)^{D/2}} \Gamma \left(\frac{D}{2}-1 \right) \left( \frac{-z^2}{4} \right)^{2-D/2} 2 \mathcal{M}^{0} (\nu) \int_0^1 dt \; t^{3-D}  \nonumber \\ = - \frac{g^2 C_F}{(4 \pi)^{D/2}} \frac{1}{\left( \frac{D}{2} - 2 \right)} \Gamma \left(\frac{D}{2}-1 \right) \left( \frac{-z^2}{4} \right)^{2 - D/2}  \mathcal{M}^{0} (\nu) \; , 
\end{gather}
where we kept $D \neq 4$ to regularise the divergence. Including the diagram $(a)$, we have 
\begin{equation}
     \mathcal{M}_{\rm Vertex, UV-sing.} (\nu, z^2, m^2) = - \frac{g^2 C_F}{(4 \pi)^{D/2}} \frac{2}{\left( \frac{D}{2} - 2 \right)} \Gamma \left(\frac{D}{2}-1 \right) \left( \frac{-z^2}{4} \right)^{2 -D/2}  \mathcal{M}^{0} (\nu) \; ,
\end{equation}
which is in agreement with the result in Ref.~\cite{Radyushkin:2017lvu}. \\

\noindent Now, we take the complete first term of eq.~(\ref{OffLightConeVertex}) and we subtract to it the term that we isolated previously. Setting $D=4$, after simple manipulations, we get
\begin{gather}
  \mathcal{M}_{\rm Vertex,b, UV-fin.} (\nu, z^2, m^2) =  \frac{g^2 C_F}{(4 \pi)^{D/2}} 4 p^0 e^{i \nu} \int_0^1 \frac{dt}{t} \int_0^1 d \beta  \left[ e^{- i \nu \beta t} \sqrt{-z^2} t \beta m K_1 \left( \sqrt{-z^2 t^2 \beta^2 m^2 } \right) - 1 \right] \; .
\end{gather}
We make the change of variable $u = \beta t$ and get
\begin{gather}
    \mathcal{M}_{\rm Vertex,b, UV-fin.} (\nu, z^2, m^2) =  \frac{g^2 C_F}{(4 \pi)^{D/2}} 4 p^0 e^{i \nu} \int_0^1 \frac{dt}{t^2} \int_0^t d u \left[ e^{- i \nu u} \sqrt{-z^2} u m K_1 \left( \sqrt{-z^2 u^2 m^2 } \right) - 1 \right] \; .
\end{gather}
Writing the integration domain as
\begin{equation}
    \int_0^1 dt \int_0^t du \; ... =  \int_0^1 du \int_u^1 dt \; ... \; ,
\end{equation}
we obtain
\begin{gather}
    \mathcal{M}_{\rm Vertex,b, UV-fin.} (\nu, z^2, m^2) = \frac{g^2 C_F}{(4 \pi)^{D/2}} \int_0^1 d u \left[ \frac{2 u}{1 - u} \right]_+ \mathcal{M}^0 (u \nu)  \sqrt{-z^2 (1-u)^2 m^2} K_1 \left( \sqrt{-z^2 (1-u)^2 m^2} \right) \; .
\end{gather}
Multiplying by a factor of two to include the diagram $(a)$, and relabeling $u$ as $\beta$, the total UV-finite term gives
\begin{gather}
 \mathcal{M}_{\rm Vertex, UV-fin.} (\nu, z^2, m^2) =    \frac{2 g^2 C_F}{(4 \pi)^{D/2}} \int_0^1 d \beta \left[ \frac{2 \beta}{1 - \beta} \right]_+ \mathcal{M}^0 (\beta \nu)  \sqrt{-z^2 (1-\beta)^2 m^2} K_1 \left( \sqrt{-z^2 (1-\beta)^2 m^2} \right) \; .
\end{gather}
In the $z^2 \rightarrow 0$ limit, this reproduces the result in Ref.~\cite{Radyushkin:2017lvu}. \\

\noindent The second term is the most complicated to treat. Also in this case, we can isolate a convenient term, \textit{i.e.} the one containing the logarithmic dependence on $\ln (-z^{2})$. It is obtained by setting $t = 1$ in the Bessel function of the integrand in the second term of eq.~(\ref{OffLightConeVertex}), \textit{i.e.}\footnote{This term is finite and we can set $D=4$.}
\begin{gather}
    \mathcal{M}_{\rm Vertex, b, evol.} (\nu, z^2, m^2) = \frac{g^2 C_F}{(4 \pi)^{D/2}}  \; 8 p^0 \int_0^1 dt \; \int_0^1 d \beta (-i \nu \beta)  e^{i (1-t) \nu + i \beta t \nu}  K_{0} \left( \sqrt{-z^2 (1-\beta)^2 m^2 } \right) \nonumber \\ = \frac{g^2 C_F}{(4 \pi)^{D/2}} \int_0^1 d \beta \left[ \frac{4 \beta}{ 1 - \beta } K_{0} \left( \sqrt{-z^2 (1-\beta)^2 m^2 } \right) \right]_+ \mathcal{M}^0 (\beta \nu)\,.
\end{gather}
Including the diagram $(a)$, we have
\begin{gather}
    \mathcal{M}_{\rm Vertex, evol.} (\nu, z^2, m^2) = \frac{2 g^2 C_F}{(4 \pi)^{D/2}} \int_0^1 d \beta \left[ \frac{4 \beta}{ 1 - \beta } K_{0} \left( \sqrt{-z^2 (1-\beta)^2 m^2 } \right) \right]_+ \mathcal{M}^0 (\beta \nu) \; .
\end{gather}

\noindent The residual term of the subtraction is
\begin{gather}
    \mathcal{M}_{\rm Vertex, b, IR-fin.} (\nu, z^2, m^2) = \frac{g^2 C_F}{(4 \pi)^{D/2}}  \; 8 p^0 \int_0^1 dt \; \int_0^1 d \beta (-i \nu \beta)  e^{i (1-t) \nu + i \beta t \nu} \nonumber \\ \times \left[  K_{0} \left( \sqrt{-z^2 (1-\beta)^2 t^2 m^2 } \right) - K_{0} \left( \sqrt{-z^2 (1-\beta)^2 m^2 } \right) \right] \; .
\end{gather}
This is the most tedious object to compute. We include the diagram $(a)$ and, after making some changes of variables, we find 
\begin{gather}
    \mathcal{M}_{\rm Vertex, IR-fin.} (\nu, z^2, m^2) = - \frac{2 g^2 C_F}{(4 \pi)^D} \int_0^1 d \beta \left( \frac{d}{d \beta} \mathcal{M}^{0} ( \beta \nu) \right) 4 \;\mathcal{F} (1-\beta, \sqrt{-z^2 m^2}) \; ,
    \label{Eq:PreByParts}
\end{gather}
where
\begin{gather}
     \mathcal{F} (1- \beta, \sqrt{-z^2 m^2})  \equiv \int_{1-\beta}^1 \frac{d t}{t} \left( 1 - \frac{1-\beta}{t} \right) \left[ K_{0} \left( \sqrt{-z^2 m^2} (1-\beta) \right) - K_{0} \left( \frac{ \sqrt{-z^2 m^2 } (1-\beta)}{t} \right) \right] \; .
\end{gather}
We now want to integrate by parts in such a way to obtain something proportional to $\mathcal{M}^{0} (\beta \nu)$ rather than to its derivative. When the derivative acts on $\mathcal{F}$ we have
\begin{gather}
    \mathcal{M}_{\rm Vertex, IR-fin.}^{(1)} (\nu, z^2, m^2) = - \frac{2 g^2 C_F}{(4 \pi)^{D/2}}  \int_0^1 d \beta \left[ 4 \Phi ( 1 - \beta, \sqrt{-z^2 m^2} )   \right] \mathcal{M}^{0} (\nu)  \; ,
\end{gather}
where
\begin{gather}
    \Phi ( 1-\beta, \sqrt{-z^2 m^2} )  
   \equiv \int_{1-\beta}^1 d t  \; \frac{\partial}{\partial \beta} \left[ \left( \frac{1-\beta}{t^2} - \frac{1}{t}  \right) \left( K_0 \left( \sqrt{-z^2 m^2 (1-\beta)^2 }  \right) -  K_0 \left( \frac{ \sqrt{-z^2 m^2 (1-\beta)^2 } }{t} \right) \right) \right] \; .
\end{gather}
This contribution is singular when the function $\Phi$ is expanded around $\beta = 1$, and it gives
\begin{gather}
    \Phi ( 1-\beta, \sqrt{-z^2 m^2} )  = \frac{\beta + \ln ( 1 -\beta)}{1-\beta} + \mathcal{O} ((1-\beta)^0) \; .
\end{gather}
The expectation is that the boundary term cancels the singularity. First, we observe that by construction $\mathcal{F} (1, \sqrt{-z^2 m^2}) = 0$, while
\begin{gather}
    \mathcal{F} (1 - \beta, \sqrt{-z^2 m^2}) \widesim{\beta \rightarrow 1}  \int_{0}^1 d \beta \; \frac{ \beta + \ln ( 1 -\beta) }{1-\beta} + R ( \sqrt{-z^2 m^2} ) \; , 
\end{gather}
where $R ( \sqrt{-z^2 m^2} )$ collects the constant terms, \textit{i.e.} those finite in the $\beta \rightarrow 1$ limit. The singular term is exactly what we need to cancel the singularity of the first term, indeed
\begin{gather}
   {\mathcal{M}}_{\rm Vertex, IR-fin.} (\nu, z^2, m^2) = - \frac{8 g^2 C_F}{(4 \pi)^{D/2}} \mathcal{M}^{0} (\nu)  R ( \sqrt{-z^2 m^2} )  \nonumber \\ - \frac{2 g^2 C_F}{(4 \pi)^{D/2}} \int_0^1 d \beta \left[ 4 \Phi ( 1 - \beta, \sqrt{-z^2 m^2} ) \mathcal{M}^{0} ( \beta \nu)  - \frac{ 4 (\beta + \ln ( 1 -\beta) ) }{1-\beta}   \; \mathcal{M}^{0} (\nu)  \right]   \; ,
   \label{Eq:Vertex_Final_Form}
\end{gather}
It might be interesting to calculate $R$ to reconstruct the starting integral in eq.~(\ref{Eq:PreByParts}).\footnote{
Although it is difficult to find an explicit form of $R$, it is easy to construct a numerical approximation of it to verify that eq.~(\ref{Eq:PreByParts}) and eq.~(\ref{Eq:Vertex_Final_Form}) are equivalent.} However, in the pseudo-distribution approach, its exact form is unnecessary. The reason is that it is proportional to $\mathcal{M}^{0} (\nu)$ and therefore vanishes at the level of the reduced Ioffe-time distribution. In this case, it is sufficient to include the singular part of $\mathcal{F}$ in order to obtain the contribution proportional to $\mathcal{M}^0 (\beta \nu)$ and the correct subtraction term. \\

\noindent Observing that $R = \mathcal{O} (z^2 m^2)$, we immediately see that the leading term in the $z^2 m^2  \rightarrow 0$ expansion gives 
\begin{gather}
    \mathcal{M}_{\rm Vertex, IR-fin.} (\nu, z^2, 0) = - \frac{2 g^2 C_F}{(4 \pi)^{D/2}} \int_0^1 d \beta  \left[ \frac{4 \ln ( 1 -\beta)}{1-\beta} + \frac{4\beta}{(1-\beta)} \right]_+ \mathcal{M}^{0} (\beta \nu) \; ,
\end{gather}
which is the correct result in the massless limit~\cite{Radyushkin:2017lvu}. \\

\noindent The complete result for the vertex contribution finally reads  
\begin{gather}
    \mathcal{M}_{\rm Vertex} (\nu, z^2, m^2) = - \frac{g^2 C_F}{(4 \pi)^{D/2}} \frac{2}{\left( \frac{D}{2} - 2 \right)} \Gamma \left(\frac{D}{2}-1 \right) \left( \frac{-z^2}{4} \right)^{2 - D/2}  \mathcal{M}^{0} (\nu) \nonumber \\ + \frac{2 g^2 C_F}{(4 \pi)^{D/2}} \int_0^1 d \beta \left[ \frac{2 \beta}{1 - \beta} \right]_+ \mathcal{M}^0 (\beta \nu) \sqrt{-z^2 (1-\beta)^2 m^2} K_1 \left( \sqrt{-z^2 (1-\beta)^2 m^2} \right) \nonumber \\ + \frac{2 g^2 C_F}{(4 \pi)^{D/2}} \int_0^1 d \beta \left[ \frac{4 \beta}{ 1 - \beta } K_{0} \left( \sqrt{-z^2 (1-\beta)^2 m^2 } \right) \right]_+ \mathcal{M}^0 (\beta \nu) - \frac{8 g^2 C_F}{(4 \pi)^{D/2}} \mathcal{M}^{0} (\nu) R ( \sqrt{-z^2 m^2} ) \nonumber \\ - \frac{ 2 g^2 C_F}{(4 \pi)^{D/2}} \int_0^1 d \beta \left[ 4 \Phi ( 1 - \beta, \sqrt{-z^2 m^2} ) \mathcal{M}^{0} ( \beta \nu)  - 4 \left( \frac{ \ln ( 1 -\beta) + \beta}{1-\beta}  \right) \; \mathcal{M}^{0} (\nu)  \right]  \; .
\end{gather}

\twocolumn

\bibliographystyle{unsrt} 
\bibliography{references}

\end{document}